\documentclass{article}

\usepackage{PRIMEarxiv}

\usepackage{algorithm} 
\usepackage{algorithmic}
\usepackage{float}

\usepackage{xcolor} 
\usepackage{amsthm}
\newtheorem{theorem}{\textcolor{black}{Theorem}}
\newcommand{\thmref}[1]{Theorem~\ref{#1}}
\newtheorem{lemma}{\textcolor{black}{Lemma}}
\newcommand{\lemref}[1]{Lemma~\ref{#1}}
\newtheorem{claim}{\textcolor{black}{Claim}}
\newcommand{\claimref}[1]{Claim~\ref{#1}}

\usepackage[numbers,square,sort]{natbib}

\usepackage{subcaption}
\usepackage{booktabs}              
\usepackage{graphicx}              
\graphicspath{{media/}}            

\usepackage{amsmath}
\usepackage{amsfonts}              
\usepackage{bm}

\usepackage[utf8]{inputenc}        
\usepackage[T1]{fontenc}           
\usepackage{microtype}             
\usepackage{url}                   

\makeatletter
\renewcommand{\eqref}[1]{Eq.\,\textup{(\ref{#1})}}
\makeatother

\usepackage{fancyhdr}

\title{Robust and Sparse Personalized Federated Learning via $\gamma$-Divergence}

\author{
  Kotaro Akimaru \\
  Graduate School of Life and Medical Sciences \\
  Doshisha University \\
  Kyoto, Japan \\
  \texttt{akimaru.kotaro@gmail.com}
  \And
  Kensuke Tanioka \\
  Department of Biomedical Sciences and Informatics \\
  Doshisha University \\
  Kyoto, Japan \\
  \texttt{ktanioka@mail.doshisha.ac.jp}
}

\begin{document}
\maketitle

\begin{abstract}
To address the challenges of statistical heterogeneity across clients and within-client data contamination in federated learning, we propose a robust and sparse personalized federated learning method based on the $\gamma$-divergence, termed $\gamma$-PFL. Specifically, we combine a loss function based on the $\gamma$-divergence with fusion and sparse regularization terms to exploit similarities in regression coefficients across clients while enabling robust estimation of personalized models and variable selection. Furthermore, to efficiently solve the estimation problem in the federated server--client system, we develop an optimization algorithm that combines the Alternating Direction Method of Multipliers with the Majorization--Minimization algorithm. Numerical simulations showed that $\gamma$-PFL maintained relatively stable estimation accuracy as the contamination proportion or outlier magnitude increased, demonstrating its robustness to within-client data contamination. In addition, a real data analysis showed that $\gamma$-PFL achieved predictive performance comparable to or better than that of an existing personalized federated learning method.
\end{abstract}


\section{Introduction}
Big data have become a key driving force behind recent advances in AI.
However, even though the total volume of data is enormous, such data are often distributed across numerous devices and organizations, such as mobile phones, medical facilities, wireless sensors, and autonomous vehicles, hereafter referred to as clients \cite{liu2025robust}.
In such distributed environments, growing privacy and security requirements have made it increasingly difficult to directly pool data held by different organizations \cite{liu2022distributed}.
Against this background, federated learning (FL) has attracted considerable attention as a framework for collaboratively training a machine learning model across multiple parties in distributed environments while addressing privacy and security concerns \cite{mcmahan2017communication,kairouz2021advances}.
In FL, each client transmits only locally computed model updates, such as model parameters or gradients, to the central server. Since each client performs local training while retaining its data locally, the raw data need not be transmitted to the central server, thereby providing a basic level of data privacy \cite{mcmahan2017communication}.
However, differences in the data-generating mechanisms across clients give rise to the challenge of statistical heterogeneity in FL \cite{mcmahan2017communication,khaled2020tighter,liu2025robust}.
In addition, real-world data may contain observations arising from data-generating mechanisms different from that underlying the target distribution. Such situations can be represented by a contamination model in which a certain proportion of observations are generated from a contaminating distribution \cite{huber2009robust}.
Therefore, it is important for FL methods to account for both statistical heterogeneity across clients and data contamination within individual clients.
Motivated by these considerations, we propose an FL method that accounts for statistical heterogeneity across clients while being robust to data contamination within each client.

To address statistical heterogeneity, various approaches have been proposed, including feature normalization \cite{huang2017arbitrary,choi2021robustnet}, model-weight regularization \cite{karimireddy2020scaffold}, and improved aggregation methods \cite{reddi2020adaptive,wang2020tackling}.
However, these methods assume a global model shared by all clients. It has been reported that, in environments with heterogeneous data-generating mechanisms, switching from locally learned models to a shared global model can lead to performance degradation \cite{yu2020salvaging,liu2025robust}.
Accordingly, personalized federated learning (PFL), which aims to learn a personalized model for each client, has attracted considerable attention \cite{smith2017federated,jiang2019improving,fallah2020personalized}.
Various approaches to PFL have been proposed, including fine-tuning \cite{fallah2020personalized}, multi-task learning (MTL) \cite{smith2017federated,huang2021personalized}, clustered federated learning \cite{ghosh2020efficient,sattler2020clustered}, parameter decoupling \cite{arivazhagan2019federated}, and knowledge distillation \cite{li2019fedmd}.
Among these approaches, we focus on utilizing similarities among local models. Specifically, following the clustered lasso formulation \cite{she2010sparse}, we impose a penalty on pairwise differences in regression coefficients corresponding to the same covariate across clients. This allows us to estimate personalized models for individual clients while capturing the cluster structure of each coefficient.
This idea, which combines elements of MTL and clustered federated learning, has been employed in Personalized Federated Learning for Robust and Sparse Regression (PerFL-RSR) \cite{liu2025robust} and Personalized Federated Learning Quantile Regression (PFLQR) \cite{lu2026personalized}, which extends PerFL-RSR.

We next consider the case in which data within individual clients are contaminated, requiring robust estimation to mitigate the effects of contamination.
However, maximum likelihood estimation and least squares estimation are known to be sensitive to outliers. To address this issue, various robust estimation methods based on M-estimation have been proposed \cite{huber2009robust}.
Maximum likelihood estimation can also be interpreted as minimizing an empirical estimate of the Kullback--Leibler divergence \cite{kawashima2023robust}. As an extension of this concept, various robust estimation methods have been proposed based on the minimization of empirical estimates of alternative divergences.
Representative examples include the density power divergence \cite{basu1998robust}, the $L_2$-divergence \cite{scott2001parametric}, and the $\gamma$-divergence \cite{fujisawa2008robust}.
In this study, we consider a setting in which the distribution of observations within each client is modeled as a mixture of the target distribution and an outlier distribution.
Among these robust estimation methods, estimators based on the $\gamma$-divergence are known to possess strong robustness: when the overlap between the outlier distribution and the target distribution is sufficiently small, the latent bias can remain sufficiently small even under heavy contamination \cite{fujisawa2008robust}.
In contrast, other robust estimation methods, including those based on the density power divergence, do not generally possess this property, and the resulting estimators may be affected by the contamination proportion \cite{kawashima2017robust}.
Accordingly, in this study, we adopt estimation based on the $\gamma$-divergence to achieve robustness against within-client contamination.

In this work, we focus on PFL in the presence of data contamination.
Furthermore, we consider high-dimensional regression problems, which are also common in FL systems with massive data sources \cite{liu2025robust}.
To address such high dimensionality, it is necessary to recover model sparsity.
Based on the above considerations, the objective of this study is to develop a robust and sparse PFL method based on the $\gamma$-divergence that simultaneously accounts for statistical heterogeneity across clients, within-client data contamination, and model sparsity. The main contributions of this study are summarized as follows. 
\begin{itemize}
    \item We propose an objective function for robust and sparse PFL based on the $\gamma$-divergence. Specifically, building on the framework of PerFL-RSR~\cite{liu2025robust}, we combine a loss function based on the $\gamma$-divergence with a fusion regularization term for personalization and a sparse regularization term for sparsity recovery. This formulation enables robust estimation against within-client data contamination while accounting for statistical heterogeneity across clients and model sparsity.

    \item To efficiently optimize the proposed objective function in the federated server--client system, we develop an optimization algorithm that combines the Alternating Direction Method of Multipliers (ADMM) \cite{boyd2011distributed} with the Majorization--Minimization (MM) algorithm \cite{hunter2004tutorial}. We refer to the resulting method as Robust and Sparse Personalized Federated Learning via $\gamma$-Divergence ($\gamma$-PFL). In particular, by introducing a two-stage majorization scheme, the proposed algorithm avoids the large-scale matrix inversion required for updating the regression coefficients and reduces the amount of information exchanged between the clients and the server.
\end{itemize}

The remainder of this paper is organized as follows.
In Section 2, we formulate the robust regression problem in the context of PFL, and in Section 3, we present the algorithm for the proposed method.
In Section 4, we evaluate the proposed method through numerical simulations, and in Section 5, we further assess its effectiveness through real data analysis.
Finally, Section 6 concludes the paper.

\section{Problem Setup}
In this section, we formulate the optimization problem for PFL while accounting for statistical heterogeneity across clients and data contamination within each client. Here, we let $[n]$ denote $\{1, \dots, n\}$.
\subsection{Model Setup}
We consider a distributed environment consisting of $M$ clients, where client $m \in [M]$ possesses a local dataset $\mathcal{D}_m = \{(\bm{x}_{mi}, y_{mi})\}_{i = 1}^{n_m}$. Here, $y_{mi} \in \mathbb{R}$ denotes the response variable and $\bm{x}_{mi} \in \mathbb{R}^{p}$ denotes the $p$-dimensional covariate vector for the $i$th observation of client $m$. Moreover, $n_m$ denotes the sample size for client $m$, and $p$ denotes the dimension of the covariates.

For client $m$, we assume the linear regression model: $y_{mi} = \bm{x}_{mi}^\top\bm{\beta}_m^{*} + \varepsilon_{mi}$, $\forall i \in [n_m]$.
The vector $\bm{\beta}_m^{*} = (\beta_{m1}^{*}, \dots, \beta_{mp}^{*})^\top$ denotes the underlying regression coefficient vector for client $m$ and is allowed to differ across clients. Also, $\varepsilon_{mi}$ denotes the error term. Here, we assume that, for each client, the error distribution is contaminated by an outlier distribution. Specifically, we consider the following contamination model for the error term, which is standard in robust statistics:
\begin{align}
  g(\varepsilon)=(1-\epsilon)f(\varepsilon)+\epsilon\delta(\varepsilon).
  \label{eq:contamination_model}
\end{align}
Here, $\varepsilon$ denotes a generic error variable, $f(\varepsilon)$ denotes the target distribution of interest, $\delta(\varepsilon)$ denotes the contamination probability density function related to outliers, and $\epsilon \in [0, 1]$ is the contamination proportion. For example, $f(\varepsilon)$ may be assumed to follow a Gaussian distribution with mean zero and variance $\sigma^2$. In contrast, $\delta(\varepsilon)$ represents a distribution generating extreme outliers in the tails of $f(\varepsilon)$ and is often modeled as a point mass.

\subsection{Client-Specific Regression Model Based on $\gamma$-Divergence}
We consider robust estimation under the contaminated distribution in \eqref{eq:contamination_model}. The difference between the estimator obtained under $g(\varepsilon)$ and that obtained under $f(\varepsilon)$ is called the latent bias, in the sense that it arises from the latent contamination factor $\delta(\varepsilon)$ generating outliers. Fujisawa and Eguchi~\cite{fujisawa2008robust} showed that estimation based on the $\gamma$-divergence can keep the latent bias sufficiently small even in the case of heavy contamination, that is, even when the contamination proportion $\epsilon$ is not small. Accordingly, in this work, we adopt a loss function based on the $\gamma$-divergence for estimating the regression model for each client. Below, we introduce the $\gamma$-divergence-based loss function defined for each client.

First, for two conditional probability density functions $g_{y\mid x}(y \mid x)$ and $f_{y\mid x}(y \mid x)$, the $\gamma$-divergence on $g_x(x)$, denoted by $D_\gamma(g_{y\mid x}, f_{y\mid x}; g_x)$, is expressed in terms of the $\gamma$-cross entropy $d_\gamma(g_{y\mid x}, f_{y\mid x}; g_x)$ as follows:
\begin{align}
  D_\gamma(g_{y\mid x}, f_{y\mid x}; g_x) = d_\gamma(g_{y\mid x}, f_{y\mid x}; g_x)
  - d_\gamma(g_{y\mid x}, g_{y\mid x}; g_x).\notag
\end{align}
Here, $\gamma > 0$ is a tuning parameter that controls the degree of robustness. If $\gamma$ is large, the method becomes more robust against contamination at the cost of efficiency, whereas if $\gamma$ is small, robustness is weakened but efficiency is recovered \cite{fujisawa2008robust}. Let $f_{y\mid x}(y \mid x; \theta)$ denote a parametric conditional probability density function corresponding to $f_{y\mid x}(y \mid x)$. Since $d_\gamma(g_{y\mid x}, g_{y\mid x}; g_x)$ is a constant that does not depend on the parameter $\theta$, estimation of $\theta$ based on minimizing $D_\gamma(g_{y\mid x}, f_{y\mid x}; g_x)$ is equivalent to estimation of $\theta$ by minimizing $d_\gamma(g_{y\mid x}, f_{y\mid x}; g_x)$. The $\gamma$-cross entropy $d_\gamma(g_{y\mid x}, f_{y\mid x}; g_x)$ is given by
\begin{align}
  d_\gamma(g_{y\mid x}, f_{y\mid x}; g_x)&=-\frac{1}{\gamma}\log\int \int
  g_{y\mid x}(y \mid x)f_{y\mid x}(y \mid x)^{\gamma}dyg_x(x)dx\notag \\
  &+\frac{1}{1+\gamma}\log\int \int f_{y\mid x}(y \mid x)^{1+\gamma}dy g_x(x) dx .
  \label{eq:gamma_cross_entropy}
\end{align}
Using the relation $g(x, y) = g_{y\mid x}(y \mid x)g_x(x)$, the first term in \eqref{eq:gamma_cross_entropy} can be written as the expectation of $f_{y\mid x}(y \mid x)^\gamma$ with respect to the joint distribution $g(x, y)$.
Let $(x_i, y_i)_{i = 1}^{n}$ be the observations independently drawn from the underlying distribution $g(x, y)$. Here, $f_{y\mid x}(y \mid x; \theta)$ denotes a parametric conditional probability density function corresponding to $f_{y\mid x}(y \mid x)$. Since $g(x, y)$ is unknown, we consider minimizing the following loss function based on the observed data:
\begin{align}
  L(\theta) = -\frac{1}{\gamma}\log \left(
  \frac{1}{n}\sum_{i=1}^n f_{y\mid x}(y_i \mid x_i; \theta)^\gamma\right)
  + \frac{1}{1+\gamma}\log\left(
  \frac{1}{n}\sum_{i=1}^n \int f_{y\mid x}(y \mid x_i; \theta)^{1+\gamma}dy\right).
  \label{eq:gamma_loss}
\end{align}
It is known that the estimator obtained by minimizing the loss function in \eqref{eq:gamma_loss} is a consistent estimator of the parameter that minimizes the $\gamma$-cross entropy $d_\gamma(g_{y\mid x}, f_{y\mid x}; g_x)$ \cite{kawashima2017robust}. Therefore, the estimator for regression analysis based on the $\gamma$-divergence is defined as
\begin{align}
  \hat{\theta} = \mathop{\arg\min}_{\theta} L(\theta).\notag
\end{align}
Here, in this work, we assume that the parametric conditional density $f_{y\mid x}(y_{mi} \mid \bm{x}_{mi}; \bm{\theta}_m)$ follows a normal distribution. In this case, for client $m$, let $\bm{\beta}_m \in \mathbb{R}^{p}$ and $\sigma_m^2 \in \mathbb{R}_{>0}$ denote the regression coefficient vector and the variance parameter, respectively, and define $\bm{\theta}_m = (\bm{\beta}_m, \sigma_m^2)$. Accordingly, we define the client-specific loss function based on the $\gamma$-divergence, denoted by $L_m(\bm{\theta}_m)$, as
\begin{align}
  L_m(\bm\theta_m) = -\frac{1}{\gamma}\log\!\Bigg\{\frac{1}{n_m}\sum_{i=1}^{n_m}
  f_{y\mid x}\!\big(y_{mi}\mid \bm{x}_{mi};\,\bm{\theta}_m\big)^{\gamma}\Bigg\}
  + \frac{1}{1+\gamma}\log\!\Bigg\{\frac{1}{n_m}\sum_{i=1}^{n_m} \int
  f_{y\mid x}\!\big(y\mid \bm{x}_{mi};\,\bm{\theta}_m\big)^{1+\gamma}\,dy\Bigg\}.
  \label{eq:client_loss}
\end{align}
In this work, our main interest is in estimating $\bm{\beta}_m$, and thus we treat $\sigma_m^2$ as a nuisance parameter. Let $\bm{\beta}=(\bm{\beta}_1^\top, \dots, \bm{\beta}_M^\top)^\top \in \mathbb{R}^{pM}$. Therefore, the optimization problem based on the average of the client-specific loss functions is formulated as
\begin{align}
  \mathop{\arg\min}_{\bm{\beta} \in \mathbb{R}^{pM}} \frac{1}{M}\sum_{m=1}^M\Bigg[
  -\frac{1}{\gamma}&\log\!\Bigg\{\frac{1}{n_m}\sum_{i=1}^{n_m}
  f_{y\mid x}\!\big(y_{mi}\mid \bm{x}_{mi};\,\bm{\theta}_m\big)^{\gamma}\Bigg\} \notag\\
  &+ \frac{1}{1+\gamma}\log\!\Bigg\{\frac{1}{n_m}\sum_{i=1}^{n_m} \int
  f_{y\mid x}\!\big(y\mid \bm{x}_{mi};\,\bm{\theta}_m\big)^{1+\gamma}\,dy\Bigg\}\Bigg].\notag
\end{align}

\subsection{Problem Formulation}
In addition to addressing data contamination within each client as described in the previous subsection, in this work, we account for statistical heterogeneity across clients by introducing a pairwise regularization term on the differences in regression coefficients across clients for each covariate. Furthermore, to address the high dimensionality issue, we impose a sparse regularization term on the regression coefficients of each client to achieve sparsity recovery.
To summarize, in this work, we achieve PFL that simultaneously accounts for statistical heterogeneity across clients and data contamination within each client by solving
\begin{align}
  \hat{\bm{\beta}}=\mathop{\arg\min}_{\bm{\beta} \in \mathbb{R}^{pM}} \; \frac{1}{M}
  &\sum_{m=1}^{M} \Bigg[
  -\frac{1}{\gamma}\log\!\Bigg\{\frac{1}{n_m}\sum_{i=1}^{n_m}
  f_{y\mid x}\!\big(y_{mi}\mid \bm{x}_{mi};\,\bm{\theta}_m\big)^{\gamma}\Bigg\} \notag \\
  &+ \frac{1}{1+\gamma}\log\!\Bigg\{\frac{1}{n_m}\sum_{i=1}^{n_m} \int
  f_{y\mid x}\!\big(y\mid \bm{x}_{mi};\,\bm{\theta}_m\big)^{1+\gamma}\,dy\Bigg\}
  \Bigg] \notag \\
  & \quad + \sum_{j=1}^{p} \sum_{m \le m'}
  p_{\lambda_1}\!\left( |\beta_{mj} - \beta_{m'j}| \right)
  + \sum_{j=1}^{p} \sum_{m=1}^{M} p_{\lambda_2}\!\left( |\beta_{mj}| \right).
  \label{eq:proposed_objective}
\end{align}
The first term in \eqref{eq:proposed_objective} is derived from the client-specific loss function based on the $\gamma$-divergence; the second term penalizes the differences in regression coefficients across clients; and the third term imposes sparse regularization on the regression coefficients for each client. Here, \(p_\lambda(\cdot)\) is a regularization function, and \(\bm{\lambda}=(\lambda_1, \lambda_2)\) denotes the regularization parameter. The parameter \(\lambda_1 > 0\) is the regularization parameter for the differences in regression coefficients across clients, and \(\lambda_2 > 0\) is the sparsity regularization parameter for the regression coefficients of each client. In this work, we consider regularization functions based on Lasso \cite{tibshirani1996regression} and SCAD \cite{fan2001variable}. Here, $\lambda > 0$ is a regularization parameter and $\iota > 2$ is the shape parameter of SCAD; the corresponding regularization functions are defined as follows:
\begin{align}
  p_\lambda^{\mathrm{Lasso}}(x)=\lambda|x|,\qquad
  p_\lambda^{\mathrm{SCAD}}(x)=
  \begin{cases}
  \lambda |x| & (|x| \le \lambda)\\
  \dfrac{-|x|^2 + 2\iota\lambda |x| - \lambda^2}{2(\iota-1)}
  & (\lambda < |x| \le \iota\lambda)\\
  \dfrac{(\iota+1)\lambda^2}{2} & (\iota\lambda < |x|).
  \end{cases}\notag
\end{align}
The proposed estimator $\hat{\bm{\beta}}$ obtained from \eqref{eq:proposed_objective} consists of $M$ personalized regression models that integrate information from all clients. Each client-specific model is robust to contaminated data containing outliers through estimation based on the $\gamma$-divergence. In addition, the proposed estimator has a group structure across clients for each covariate, which enables the analysis of similarity among clients. Moreover, the sparse regularization term addresses the high dimensionality issue by enabling variable selection and sparsity recovery in high-dimensional settings, thereby yielding interpretable models with reduced overfitting.

\section{Proposed Method}\label{sec:proposed_method}
In this section, we derive the update rules for each parameter in the proposed method. We also present the proposed algorithm based on these update rules.

\paragraph{Notation:}
The notation $\otimes$ denotes the Kronecker product. For any vector $\bm{x}$, we use $\|\bm{x}\|_2 = (\bm{x}^\top\bm{x})^{1/2}$ to denote the $\ell_2$-norm. For any positive definite matrix $\bm{H}$, we define $\|\bm{x}\|_{\bm{H}}^2 = \bm{x}^\top\bm{H}\bm{x}$. Moreover, $\mathrm{diag}(x_1, \dots, x_n)$ denotes the diagonal matrix with diagonal entries $x_1, \dots, x_n$, and $\mathrm{blkdiag}(\bm{X}_1, \dots, \bm{X}_N)$ denotes the block diagonal matrix with diagonal blocks $\bm{X}_1, \dots, \bm{X}_N$. In addition, $\nabla f(\bm{\beta})$ denotes the gradient of a function $f$ with respect to $\bm{\beta}$, and $\varsigma_{\max}(\bm{X})$ denotes the largest eigenvalue of a symmetric matrix $\bm{X}$. Furthermore, $\operatorname{sgn}(\cdot)$ denotes the sign function, and $I[\cdot]$ denotes the indicator function.

\subsection{Problem Reformulation}
In the proposed method, the parameter $\bm{\beta}$ is estimated using ADMM \cite{boyd2011distributed}. To this end, \eqref{eq:proposed_objective} is decomposed into $f_\gamma(\bm{\beta})$ and $h(\bm{\delta})$, where the latter is defined by introducing an auxiliary variable $\bm{\delta}$. This decomposition leads to a constrained optimization problem, based on which the augmented Lagrangian is defined.

\paragraph{Definition of $f_\gamma(\bm{\beta})$:}
The first term in \eqref{eq:proposed_objective} is defined as $f_\gamma(\bm{\beta})$ as follows:
\begin{align}
  f_{\gamma}(\bm{\beta}) =\frac{1}{M} \sum_{m=1}^{M} \Bigg[
  -\frac{1}{\gamma}\log\!\Bigg\{\frac{1}{n_m}\sum_{i=1}^{n_m}
  f_{y\mid x}\!\big(y_{mi}\mid \bm{x}_{mi};\,\bm{\theta}_m\big)^{\gamma}\Bigg\}
  +\frac{1}{1+\gamma}\log\!\Bigg\{\frac{1}{n_m}\sum_{i=1}^{n_m} \int
  f_{y\mid x}\!\big(y\mid \bm{x}_{mi};\,\bm{\theta}_m\big)^{1+\gamma}\,dy\Bigg\} \Bigg].\notag
\end{align}

\paragraph{Definition of $h(\bm{\delta})$:}
We express the second and third terms in \eqref{eq:proposed_objective} by a function $h(\bm{\delta})$ depending on an auxiliary variable $\bm{\delta}\in \mathbb{R}^{p\frac{M(M-1)}{2}+pM}$.
Specifically, we partition $\bm{\delta}$ as $\bm{\delta}=\bigl(\bm{\delta}_{\mathrm{fused}}^\top,\bm{\delta}_{\mathrm{sparse}}^\top\bigr)^\top$, where $\bm{\delta}_{\mathrm{fused}}\in \mathbb{R}^{p\frac{M(M-1)}{2}}$ corresponds to the fused penalty and $\bm{\delta}_{\mathrm{sparse}}\in \mathbb{R}^{pM}$ corresponds to the sparsity penalty.
We also define the index sets $\mathcal{J}_{\mathrm{fused}} = \{1, 2, \dots, pM(M - 1)/2\}$ and $\mathcal{J}_{\mathrm{sparse}} = \{pM(M - 1)/2 + 1, \dots, pM(M - 1)/2 + pM\}$.
For $j\in\mathcal{J}_{\mathrm{fused}}$, $\delta_j$ is a component of $\bm{\delta}_{\mathrm{fused}}$, whereas for $j\in\mathcal{J}_{\mathrm{sparse}}$, $\delta_j$ is a component of $\bm{\delta}_{\mathrm{sparse}}$.
Furthermore, let $\bm{e}_k$ denote the standard basis vector whose $k$th component is equal to $1$, and let $\bm{I}_{pM}$ denote the $pM\times pM$ identity matrix.
We define $\bm{E}=(\bm{e}_i-\bm{e}_j,\  i<j)^\top$ and $\bm{\Omega}=\bm{E}\otimes\bm{I}_p$.
We define $\bm{\delta}_{\mathrm{fused}}=\bm{\Omega}\bm{\beta}$ and $\bm{\delta}_{\mathrm{sparse}}=\bm{\beta}=\bm{I}_{pM}\bm{\beta}$.
With $\bm{A}=[\bm{\Omega}^{\top},\bm{I}_{pM}]^\top$, $\bm{\delta}$ can be written as
\begin{align}
  \bm{\delta}
  = \begin{pmatrix}
  \bm{\delta}_{\text{fused}} \\ \bm{\delta}_{\text{sparse}}
  \end{pmatrix}
  = \begin{pmatrix}
  \bm{\Omega}\bm{\beta} \\ \bm{I}_{pM}\bm{\beta}
  \end{pmatrix}
  = \begin{pmatrix}
  \bm{\Omega} \\ \bm{I}_{pM}
  \end{pmatrix} \bm{\beta}
  = \bm{A}\bm{\beta}.\notag
\end{align}
Therefore, by imposing the constraint $\bm{A}\bm{\beta}-\bm{\delta}=\bm{0}$, the fused-penalty term and the sparsity-penalty term can be represented using the auxiliary variable $\bm{\delta}$.
Accordingly, the second and third terms in \eqref{eq:proposed_objective} are defined as
\begin{align}
  h(\bm{\delta})&=\sum_{j\in\mathcal{J}_{\mathrm{fused}}} p_{\lambda_1}\!\left( |\delta_j| \right)
  +\sum_{j\in\mathcal{J}_{\mathrm{sparse}}} p_{\lambda_2}\!\left( |\delta_j| \right).\notag
\end{align}

\paragraph{Augmented Lagrangian:}
With the above definitions, the optimization problem in \eqref{eq:proposed_objective} can be reformulated as the following constrained optimization problem:
\begin{equation}
  \min \; f_\gamma(\bm{\beta}) + h(\bm{\delta}), \quad \text{s.t.} \quad \bm{A}\bm{\beta} = \bm{\delta}.\notag
\end{equation}
For this constrained optimization problem, the augmented Lagrangian function $L_\rho(\bm{\beta},\bm{\delta},\bm{u})$ is defined, with Lagrange multiplier $\bm{u} \in \mathbb{R}^{p\frac{M(M-1)}{2}+pM}$ and penalty parameter $\rho \in \mathbb{R}_{>0}$, as follows:
\begin{align}
  L_\rho(\bm{\beta},\bm{\delta},\bm{u}) =& f_\gamma(\bm{\beta}) + h(\bm{\delta})
  + \bm{u}^\top(\bm{A}\bm{\beta} - \bm{\delta}) + \frac{\rho}{2} \|\bm{A}\bm{\beta} - \bm{\delta}\|_2^2.\notag
\end{align}
Specifically, based on ADMM, the update rules for $\bm{\beta}$, $\bm{\delta}$, and $\bm{u}$ at the $t$th iteration are given by
\begin{align}
  \bm{\beta}^{(t+1)} &\leftarrow \mathop{\arg\min}_{\bm{\beta}}\,
  L_\rho(\bm{\beta},\bm{\delta}^{(t)},\bm{u}^{(t)}),\notag\\
  \bm{\delta}^{(t+1)} &\leftarrow \mathop{\arg\min}_{\bm{\delta}}\, L_\rho(\bm{\beta}^{(t+1)}, \bm{\delta},
  \bm{u}^{(t)}),
  \label{eq:admm_delta_update} \\
  \bm{u}^{(t+1)} &\leftarrow \bm{u}^{(t)} + \rho(\bm{A}\bm{\beta}^{(t+1)} - \bm{\delta}^{(t+1)}).
  \label{eq:admm_dual_update}
\end{align}

\subsection{Parameter Updates}
Here, we derive the update rules for the parameters in the proposed method. The details of whether each update is performed on the server or on the clients are provided in subsection 3.3.
\subsubsection{Update of $\bm{\beta}$}
To derive the update rule for $\bm{\beta}$, directly minimizing $f_\gamma(\bm{\beta})$ is not straightforward. We therefore first construct a surrogate function $F_\gamma(\bm{\beta})$ based on the MM algorithm \cite{hunter2004tutorial}, following the sparse $\gamma$-linear regression \cite{kawashima2017robust} framework.
We next construct a surrogate function $\tilde{F}_\gamma(\bm{\beta})$ for $F_\gamma(\bm{\beta})$ and derive the update rule for $\bm{\beta}$ in a form analogous to that of PerFL-RSR~\cite{liu2025robust}.

\paragraph{Construction of $F_\gamma(\bm{\beta})$ and its limitation:}
We begin by introducing the majorizing surrogate function $F_\gamma(\bm{\beta})$ given in Lemma~\ref{lem:majorizing_function} and consider minimizing it.

\begin{lemma}[Construction of a surrogate function]
\label{lem:majorizing_function}
By the MM algorithm, we obtain the following surrogate function for
$f_\gamma(\bm{\beta})$, where
$\bm{\sigma}^2=(\sigma_1^2, \dots, \sigma_M^2)^\top$:
\begin{align}
  F_\gamma(\bm{\beta},\bm{\sigma}^2) = \frac{1}{M} \sum_{m=1}^{M} \Bigg\{ \frac{1}{2(1+\gamma)}\log\sigma_m^2
  + \frac{1}{2\sigma_m^2} \sum_{i=1}^{n_m} w_{mi}^{(t)} \bigl\{y_{mi}-\bm{x}_{mi}^\top\bm{\beta}_{m}\bigr\}^2
  \Bigg\}.
  \label{eq:surrogate_full}
\end{align}

Here, $w_{mi}^{(t)}$ is a weight assigned to observation $i$ on client $m$, defined by
\begin{align}
  w_{mi}^{(t)} = \frac{ f_{y\mid x}\!\bigl(y_{mi}\mid \bm{x}_{mi};\,\bm{\theta}_m^{(t)}\bigr)^{\gamma} }{
  \sum_{k=1}^{n_m} f_{y\mid x}\!\bigl(y_{mk}\mid \bm{x}_{mk};\,\bm{\theta}_m^{(t)}\bigr)^{\gamma} } .
  \label{eq:weight_update}
\end{align}
\end{lemma}
For client $m$, let $\bm{X}_m=\bigl(\bm{x}_{m1}, \dots, \bm{x}_{mn_m}\bigr)^\top$, $\bm{y}_m=\bigl(y_{m1}, \dots, y_{mn_m}\bigr)^\top$, and $\bm{W}_m^{(t)}=\mathrm{diag}(w_{m1}^{(t)}, \dots, w_{mn_m}^{(t)})$. We further define $\tilde{\bm{X}}_m^{(t)}=\bigl(\bm{W}_m^{(t)}\bigr)^{1/2}\bm{X}_m/\sigma_m^{(t)}$ and $\tilde{\bm{y}}_m^{(t)}=\bigl(\bm{W}_m^{(t)}\bigr)^{1/2}\bm{y}_m/\sigma_m^{(t)}$. Using these quantities, we construct the block matrix $\tilde{\bm{X}}^{(t)}=\mathrm{blkdiag}(\tilde{\bm{X}}_1^{(t)}, \dots, \tilde{\bm{X}}_M^{(t)})$ and the stacked vector $\tilde{\bm{y}}^{(t)}=\bigl((\tilde{\bm{y}}_1^{(t)})^\top, \dots, (\tilde{\bm{y}}_M^{(t)})^\top\bigr)^\top$.
With these definitions, the part of \eqref{eq:surrogate_full} that depends on
$\bm{\beta}$ can be written as follows:
\begin{align}
  F_\gamma(\bm{\beta}) = \frac{1}{2M}\, \|\tilde{\bm{y}}^{(t)}-\tilde{\bm{X}}^{(t)}\bm{\beta}\|_2^2 .\notag
\end{align}

Accordingly, fixing $\tilde{\bm{X}}^{(t)}$ and $\tilde{\bm{y}}^{(t)}$, which are constructed from $w_{mi}^{(t)}$ and $\sigma_m^{2\,(t)}$, we consider the following minimization problem:
\begin{align}
  \bm{\beta}^{(t+1)} &=\mathop{\arg\min}_{\bm{\beta}}
  \frac{1}{2M}\bigl\|\tilde{\bm{y}}^{(t)}-\tilde{\bm{X}}^{(t)}\bm{\beta}\bigr\|_2^2
  + \bm{u}^{(t)\top}\!\bigl(\bm{A}\bm{\beta} - \bm{\delta}^{(t)}\bigr)
  + \frac{\rho}{2}\bigl\|\bm{A}\bm{\beta} - \bm{\delta}^{(t)}\bigr\|_2^2.
  \label{eq:beta_update_subproblem}
\end{align}

By differentiating \eqref{eq:beta_update_subproblem} with respect to $\bm{\beta}$ and solving the corresponding stationary condition, we obtain the following closed-form update:
\begin{align}
  \bm{\beta}^{(t+1)} &= \Bigl\{\frac{1}{M}\tilde{\bm{X}}^{(t)\top}\tilde{\bm{X}}^{(t)}
  +\rho\bm{A}^\top\bm{A}\Bigr\}^{-1} \Bigl( \frac{1}{M}\tilde{\bm{X}}^{(t)\top}\tilde{\bm{y}}^{(t)}
  -\bm{A}^\top\bm{u}^{(t)} +\rho\bm{A}^\top\bm{\delta}^{(t)} \Bigr) .
  \label{eq:beta_update_closed_form}
\end{align}

However, \eqref{eq:beta_update_closed_form} requires inversion of a $pM\times pM$ matrix. When either the number of clients $M$ or the number of covariates $p$ is large, this leads to substantial computational and memory costs. Therefore, directly using \eqref{eq:beta_update_closed_form} is not desirable from the viewpoint of computational efficiency.
\paragraph{Construction of $\tilde{F}_\gamma(\bm{\beta})$ and the resulting update:}
Let $\tau\in \mathbb{R}_{>0}$ be a tuning parameter and $\bm{H}\in \mathbb{R}^{pM\times pM}$ be a positive definite matrix. We construct the function $\tilde{F}_\gamma(\bm{\beta})$ as
\begin{align}
  \tilde{F}_\gamma(\bm{\beta}) = F_\gamma(\bm{\beta}^{(t)})
  + \nabla F_\gamma(\bm{\beta}^{(t)})^\top (\bm{\beta}-\bm{\beta}^{(t)})
  + \frac{1}{2\tau}\|\bm{\beta}-\bm{\beta}^{(t)}\|_{\bm{H}}^2 .\notag
\end{align}
Using $\tilde{F}_\gamma(\bm{\beta})$, we consider the following minimization problem:
\begin{align}
  \bm{\beta}^{(t+1)} = \mathop{\arg\min}_{\bm{\beta}} \tilde{F}_\gamma(\bm{\beta})
  + \bm{u}^{(t)\top}(\bm{A}\bm{\beta}-\bm{\delta}^{(t)})
  + \frac{\rho}{2}\|\bm{A}\bm{\beta}-\bm{\delta}^{(t)}\|_2^2.
  \label{eq:tilde_surrogate_problem}
\end{align}
By differentiating \eqref{eq:tilde_surrogate_problem} with respect to $\bm{\beta}$ and solving the corresponding stationary condition, we obtain the following closed-form update:
\begin{align}
  \bm{\beta}^{(t+1)} = \left(\frac{1}{\tau}\bm{H}+\rho\bm{A}^\top\bm{A}\right)^{-1}
  \left\{ \frac{1}{\tau}\bm{H}\bm{\beta}^{(t)} - \nabla F_\gamma(\bm{\beta}^{(t)})
  + \rho\bm{A}^\top\bm{\delta}^{(t)} - \bm{A}^\top\bm{u}^{(t)} \right\}.\notag
\end{align}
Furthermore, to avoid computing the matrix inverse in the above expression, we define $\bm{H}$ using the rank parameter $r$ as $\bm{H}=r\bm{I}-\rho\tau\bm{A}^\top\bm{A}$. Therefore, the update rule for $\bm{\beta}$ can be rewritten as
\begin{align}
  \bm{\beta}^{(t+1)}= \frac{1}{r}\bm{H}\bm{\beta}^{(t)}- \frac{\tau}{r}
  \left\{\nabla F_\gamma(\bm{\beta}^{(t)})-\rho\bm{A}^\top\bm{\delta}^{(t)}+\bm{A}^\top\bm{u}^{(t)}\right\}.
  \label{eq:beta_update_server}
\end{align}
However, for $\tilde{F}_\gamma(\bm{\beta})$ to majorize $F_\gamma(\bm{\beta})$ under the MM algorithm, \thmref{thm:rank_condition} shows that the rank parameter $r$ must satisfy \eqref{eq:rank_condition}.

\begin{theorem}[Sufficient condition for the rank parameter $r$]
\label{thm:rank_condition}
Let $\bm{H}\in\mathbb{R}^{pM\times pM}$ be a positive definite matrix, and define
$\bm{H}=r\bm{I}_{pM}-\rho\tau\bm{A}^\top\bm{A}$.
If the rank parameter $r$ satisfies
\begin{align}
  r \ge \rho\tau\,\varsigma_{\max}(\bm{A}^\top\bm{A})
  + \frac{\tau}{M} \varsigma_{\max} ( \tilde{\bm{X}}^{(t)\top}\tilde{\bm{X}}^{(t)} ),
  \label{eq:rank_condition}
\end{align}
then the following inequality holds for any $\bm{\beta}$:
\begin{align}
  F_\gamma(\bm{\beta}) \le F_\gamma(\bm{\beta}^{(t)})
  + \nabla F_\gamma(\bm{\beta}^{(t)})^\top (\bm{\beta}-\bm{\beta}^{(t)})
  + \frac{1}{2\tau} \|\bm{\beta}-\bm{\beta}^{(t)}\|_{\bm{H}}^2.
  \label{eq:majorization_bound}
\end{align}
\end{theorem}

\subsubsection{Updates of $\bm{\delta}$}
The terms in \eqref{eq:admm_delta_update} involving $\bm{\delta}$ can be rearranged as follows:
\begin{align}
  \bm{\delta}^{(t+1)}&=\mathop{\arg\min}_{\bm{\delta}} \;h(\bm{\delta})+
  \frac{\rho}{2}\|\bm{\delta}-(\bm{A}\bm{\beta}^{(t+1)}+\frac{1}{\rho}\bm{u}^{(t)})\|_2^2.
  \label{eq:delta_update_expression}
\end{align}
For \eqref{eq:delta_update_expression}, the parameter $\bm{\delta}$ is updated by applying, to each component $\delta_j$ of $\bm{\delta}$, the proximal operator $T_{\lambda,\rho}(y)$, which is defined as the solution to $\mathop{\arg\min}_{x\in\mathbb{R}}\left\{p_\lambda(x)+\frac{\rho}{2}(x-y)^2\right\}$ \cite{polson2015proximal}. In particular, the proximal operators corresponding to $p_\lambda^{\mathrm{Lasso}}(x)$ and $p_\lambda^{\mathrm{SCAD}}(x)$ are given by $T_{\lambda,\rho}^{\mathrm{Lasso}}(y)$ and $T_{\lambda,\rho}^{\mathrm{SCAD}}(y)$, respectively, as follows:
\begin{align}
  T^{\mathrm{Lasso}}_{\lambda,\rho}(y) = \mathrm{ST}(y, \lambda/\rho) ,\quad
  T^{\mathrm{SCAD}}_{\lambda,\rho}(y) =
  \begin{cases}
  \mathrm{ST}(y, \lambda/\rho) & (|y| \le \lambda + \lambda/\rho) \\
  \displaystyle
  \frac{(\iota \rho - \rho) \mathrm{ST}(y, \iota\lambda/(\iota \rho - \rho))}
  {\iota \rho - \rho - 1}
  & (\lambda + \lambda/\rho < |y| \le \iota \lambda) \ \\
  y & (|y| > \iota \lambda) .
  \end{cases}\notag
\end{align}
Here, $\mathrm{ST}(x,\lambda)\ \ (x\in\mathbb{R},\  \lambda\in\mathbb{R}_{>0})$ denotes the soft-thresholding operator \cite{donoho1995adapting}, which is defined as $\mathrm{ST}(x,\lambda)=\operatorname{sgn}(x)(|x|-\lambda)\,I[|x|-\lambda>0]$. Therefore, using $T_{\lambda,\rho}(y)$, each component of $\bm{\delta}$ is updated as
\begin{align}
  \delta_j^{(t+1)} = T_{\lambda_j,\rho}\left([\bm{A}\bm{\beta}^{(t+1)} + \rho^{-1}\bm{u}^{(t)}]_j\right).
  \label{eq:proposed_delta_update}
\end{align}
Here, $\lambda_j$ denotes the regularization parameter assigned according to the index $j$. Specifically, we set $\lambda_j=\lambda_1$ for $j\in\mathcal{J}_{\mathrm{fused}}$, and $\lambda_j=\lambda_2$ for $j\in\mathcal{J}_{\mathrm{sparse}}$.

\subsubsection{Updates of Other Parameters}
The weights $w_{mi}$ and the variance $\sigma_m^2$ also need to be updated iteratively. These updates are derived based on the same idea as the update scheme for sparse $\gamma$-linear regression proposed by Kawashima and Fujisawa~\cite{kawashima2017robust}. The corresponding update rules are given below.
\paragraph{Update of $w_{mi}$:}
The weights $w_{mi}^{(t)}$ are introduced to construct the surrogate function $F_\gamma(\bm{\beta})$ in the MM algorithm. Accordingly, their update rule is given by \eqref{eq:weight_update} in Lemma~\ref{lem:majorizing_function}.

\paragraph{Update of $\sigma_m^2$:}
For the update of $\sigma_m^2$, we consider the minimization problem $\bm{\sigma}^{2(t+1)}=\mathop{\arg\min}_{\bm{\sigma}^2} F_\gamma(\bm{\beta},\bm{\sigma}^2)$. Here, for each client $m$, we define a local function $F_{\gamma(m)}(\bm{\beta}_m,\sigma_m^2)$ such that $F_\gamma(\bm{\beta},\bm{\sigma}^2)=\frac{1}{M}\sum_{m=1}^{M}F_{\gamma(m)}(\bm{\beta}_m,\sigma_m^2)$, where
\begin{align}
  F_{\gamma(m)}(\bm{\beta}_m,\sigma_m^2) = \frac{1}{2(1+\gamma)}\log \sigma_m^2
  + \frac{1}{2\sigma_m^2}\sum_{i=1}^{n_m}
  w_{mi}^{(t)}\bigl(y_{mi}-\bm{x}_{mi}^\top\bm{\beta}_m\bigr)^2 .\notag
\end{align}
Since the terms corresponding to the other clients are constant when optimizing with respect to $\sigma_m^2$, the variance parameter can be updated independently for each client $m$. Accordingly, the update for client $m$ is given by $\sigma_m^{2\,(t+1)}=\mathop{\arg\min}_{\sigma_m^2>0}F_{\gamma(m)}(\bm{\beta}_m^{(t+1)},\sigma_m^2)$. By differentiating $F_{\gamma(m)}(\bm{\beta}_m,\sigma_m^2)$ with respect to $\sigma_m^2$ and solving the resulting stationary condition, we obtain the following closed-form update:
\begin{align}
  \sigma_m^{2\,(t+1)} &= (1+\gamma)\, \left(\bm{y}_m-\bm{X}_m\bm{\beta}_m^{(t+1)}\right)^\top \bm{W}_m^{(t)}
  \left(\bm{y}_m-\bm{X}_m\bm{\beta}_m^{(t+1)}\right).
  \label{eq:proposed_sigma_update}
\end{align}

\subsection{Proposed Algorithm}
In FL, each client collaboratively updates the model parameters while retaining its local data and communicating with the server. Here, following the idea of PerFL-RSR, we develop an update algorithm to achieve communication efficiency between the clients and the server.

First, we present the parameters updated on the server. On the server side, $\bm{\delta}$ and $\bm{u}$ are updated based on the similarity of the regression coefficients corresponding to the same covariate across clients. Since these updates require the coefficient vector $\bm{\beta}$, it must be maintained on the server.
However, it follows from \eqref{eq:beta_update_server} that updating $\bm{\beta}$ requires the computation of $\nabla F_\gamma(\bm{\beta}^{(t)})$, which involves information from the local data of each client. Therefore, $\bm{\beta}$ cannot be updated solely on the server side. To address this issue, the server computes the following quantity $\tilde{\bm{\beta}}^{(t)}$, which corresponds to the part of the update formula for $\bm{\beta}$ that does not require local data:
\begin{align}
  \tilde{\bm{\beta}}^{(t)} = \frac{1}{r}\bm{H}\bm{\beta}^{(t)} -\frac{\tau}{r}
  \left( -\rho\bm{A}^\top\bm{\delta}^{(t)} +\bm{A}^\top\bm{u}^{(t)} \right).
  \label{eq:proposed_tilde_beta}
\end{align}

Here, let $\tilde{\bm{\beta}}^{(t)}=(\tilde{\bm{\beta}}_1^{(t)\top}, \dots, \tilde{\bm{\beta}}_M^{(t)\top})^\top \in \mathbb{R}^{pM}$, where $\tilde{\bm{\beta}}_m^{(t)} \in \mathbb{R}^p$ is sent from the server to client $m$. For each client $m$, $\bm{\beta}_m$ is updated using the local gradient $\nabla F_{\gamma(m)}(\bm{\beta}_m^{(t)})$ given in Lemma~\ref{lem:local_gradient} as
\begin{align}
  \bm{\beta}_m^{(t+1)} = \tilde{\bm{\beta}}_m^{(t)} -\frac{\tau}{rM}\nabla F_{\gamma(m)}(\bm{\beta}_m^{(t)}).
  \label{eq:proposed_client_beta_update}
\end{align}

\begin{lemma}[Derivation of $\nabla F_{\gamma(m)}(\bm{\beta}_m^{(t)})$]
\label{lem:local_gradient}
Let the gradient of the surrogate function $F_\gamma(\bm{\beta})=\frac{1}{2M}\|\tilde{\bm{y}}^{(t)}-\tilde{\bm{X}}^{(t)}\bm{\beta}\|_2^2$ at $\bm{\beta}^{(t)}$ be written as
$\nabla F_\gamma(\bm{\beta}^{(t)})
= \frac{1}{M}\bigl(
\nabla F_{\gamma(1)}(\bm{\beta}_1^{(t)})^\top,
\nabla F_{\gamma(2)}(\bm{\beta}_2^{(t)})^\top,
\dots,
\nabla F_{\gamma(M)}(\bm{\beta}_M^{(t)})^\top
\bigr)^\top$.
In this case, the local gradient corresponding to client $m$, denoted by $\nabla F_{\gamma(m)}(\bm{\beta}_m^{(t)})$, is given by
\begin{align}
  \nabla F_{\gamma(m)}(\bm{\beta}_m^{(t)}) = \frac{1}{\sigma_m^{2\,(t)}}
  \left( \bm{X}_m^\top\bm{W}_m^{(t)}\bm{X}_m\bm{\beta}_m^{(t)}
  - \bm{X}_m^\top\bm{W}_m^{(t)}\bm{y}_m \right).\notag
\end{align}
\end{lemma}

Each client $m$ sends the updated $\bm{\beta}_m^{(t+1)}$ to the server. The server then concatenates them as $\bm{\beta}^{(t+1)}=(\bm{\beta}_1^{(t+1)\top}, \dots, \bm{\beta}_M^{(t+1)\top})^\top$, thereby completing the update of $\bm{\beta}$.

Next, we describe the client-side updates. In addition to $\bm{\beta}_m$, the weight $w_{mi}$ and the variance $\sigma_m^2$ are also updated using the local data on each client. As in the sparse $\gamma$-linear regression of Kawashima and Fujisawa~\cite{kawashima2017robust}, these quantities are updated in the order of $w_{mi}$, $\bm{\beta}_m$, and $\sigma_m^2$.

\begin{figure}[H]
  \centering
  \includegraphics[
    width=0.82\linewidth,
    trim=80 120 90 80, 
    clip
  ]{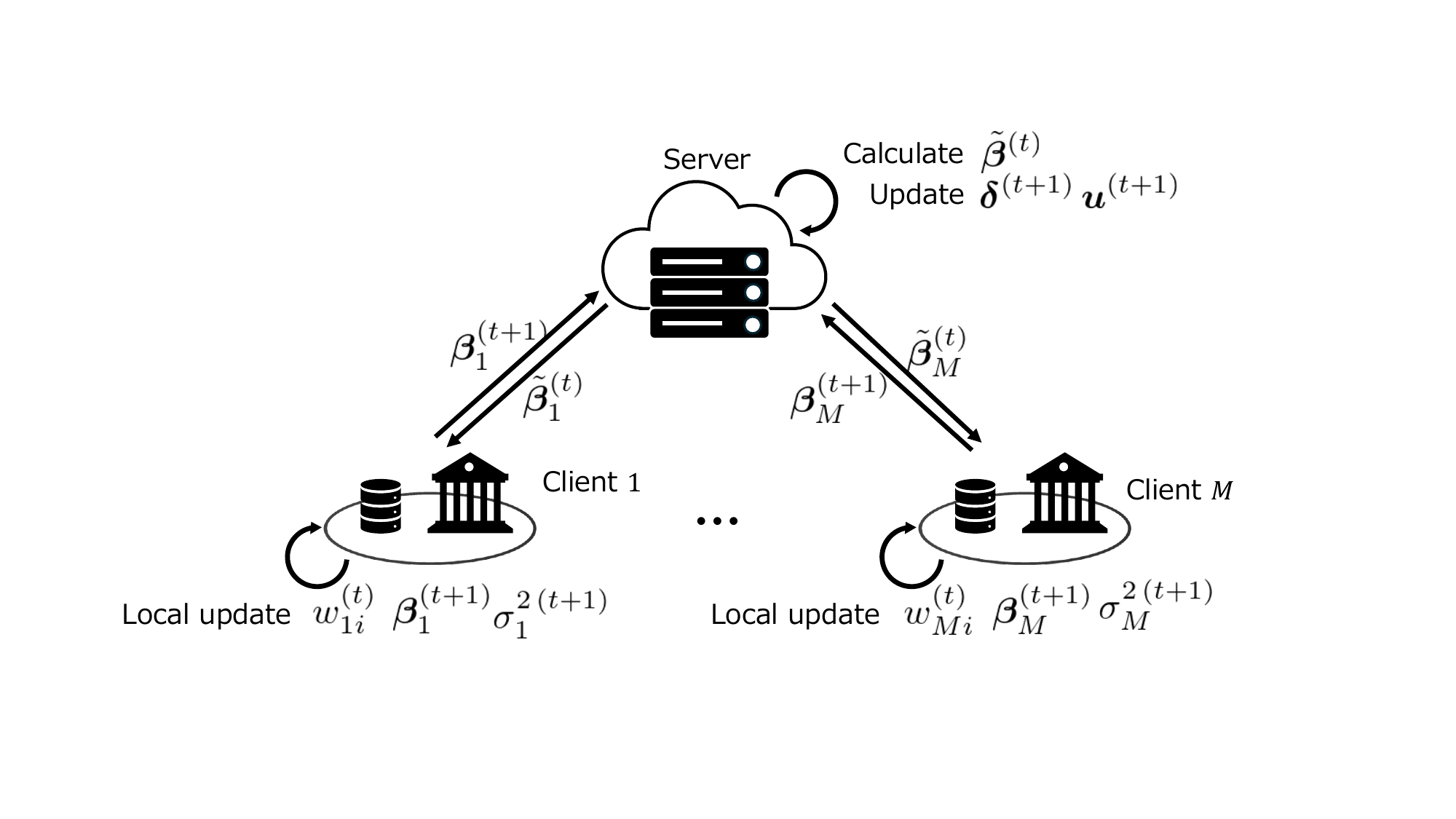}
  \caption{Illustration of $\gamma$-PFL.}
  \label{fig:communication}
\end{figure}

Accordingly, the update scheme of the proposed method is illustrated in Figure~\ref{fig:communication} and summarized in Algorithm~\ref{alg:proposed}. For the initial parameter values, we use stable estimates designed for high-dimensional settings.
Specifically, for each client $m$, $\bm{\beta}_m^{(0)}$ is estimated by fitting ridge regression~\cite{hoerl1970ridge} to the local data, and $\sigma_m^{2(0)}$ is computed from the residual sum of squares on the corresponding local dataset.

\begin{algorithm}[H]
\caption{$\gamma$-PFL Algorithm}
\label{alg:proposed}
\begin{algorithmic}[1]
\REQUIRE Tuning parameters $\tau, \rho, \gamma, \lambda_1, \lambda_2$,
and $\iota$ only when SCAD regularization is used; number of iterations $T$;
\STATE \textbf{Initialization:} Set $\bm{u}^{(0)}=\bm{0}$; compute the initial $\bm{\beta}_m^{(0)}$ and $\sigma_m^{2(0)}$ by ridge regression; upload $\bm{\beta}^{(0)}$ to the server;
\FOR{$t = 0, 1, \dots, T - 1$}
    \STATE \textbf{Server:}
    \STATE \quad Calculate $\tilde{\bm{\beta}}^{(t)}$ by \textup{(\ref{eq:proposed_tilde_beta})};
    \STATE \textbf{Clients $m \in [M]$} do in parallel:
    \STATE \quad Download $\tilde{\bm{\beta}}_m^{(t)}$ from server;
    \STATE \quad Update $w_{mi}^{(t)}$ for $i = 1, \dots, n_m$ by \textup{(\ref{eq:weight_update})};
    \STATE \quad Update $\bm{\beta}_m^{(t+1)}$ by \textup{(\ref{eq:proposed_client_beta_update})};
    \STATE \quad Update $\sigma_m^{2\,(t+1)}$ by \textup{(\ref{eq:proposed_sigma_update})};
    \STATE \quad Send $\bm{\beta}_m^{(t+1)}$ to the server;
    \STATE \textbf{Server:}
    \STATE \quad Update $\bm{\delta}^{(t+1)}$ by \textup{(\ref{eq:proposed_delta_update})};
    \STATE \quad Update $\bm{u}^{(t+1)}$ by \textup{(\ref{eq:admm_dual_update})};
\ENDFOR
\ENSURE Each client $m \in [M]$ returns $\bm{\beta}_m^{(T)}$, and the server returns $\bm{\beta}^{(T)}$.
\end{algorithmic}
\end{algorithm}

\subsection{Selection of Tuning Parameters}
Among the parameters involved in the proposed method, the algorithmic parameters $\tau$ and $\rho$ used in the optimization algorithm, together with the shape parameter $\iota$ of the SCAD penalty, are fixed at $\tau = 1$, $\rho = 2$, and $\iota = 3$, following PerFL-RSR~\cite{liu2025robust}. In contrast, the regularization parameters $\lambda_1$ and $\lambda_2$, as well as the robustness parameter $\gamma$, are selected by 5-fold cross-validation (CV)~\cite{stone1974cross}. The tuning procedure is summarized in Algorithm~\ref{alg:cv_tuning}.

For each client $m$, the local dataset $\mathcal{D}_m$ is divided into five folds. In the $k$th validation step, the $k$th fold is used as the test dataset, whereas the remaining four folds are used as the training dataset. These datasets are denoted by $\mathcal{D}_{m,\mathrm{test}}^{(k)}$ and $\mathcal{D}_{m,\mathrm{train}}^{(k)}$, respectively. Let $\Gamma$, $\Lambda_1$, and $\Lambda_2$ denote the sets of candidate values for $\gamma$, $\lambda_1$, and $\lambda_2$, respectively.
For $\Gamma$, to examine a broader range of robustness levels, we extended the values used by Kawashima and Fujisawa~\cite{kawashima2017robust} and set $\Gamma = \{0.1, 0.3, 0.5, 0.7, 0.9\}$.
For $\Lambda_1$ and $\Lambda_2$, the search ranges were determined for each simulation setting to include values for which most of the estimated coefficients were nonzero as well as values for which most were zero.
We define the set of all candidate combinations as
$\mathcal{T}=\Gamma\times\Lambda_1\times\Lambda_2.$
The predictive performance in each fold is evaluated using the median absolute residual (MAR). For a candidate triplet $(\gamma,\lambda_1,\lambda_2)\in\mathcal{T}$, when the $k$th fold is used as the test dataset, the MAR for client $m$ is defined as
$\mathrm{MAR}_{m}^{(k)}(\gamma,\lambda_1,\lambda_2)=\mathrm{median}(\{|y_{mi}-\bm{x}_{mi}^\top\hat{\bm{\beta}}_{m}^{(k)}|:i = 1, \dots, n_{m,k}\}).$
The overall MAR for the $k$th fold is then defined by taking the median across clients:
$\mathrm{MAR}^{(k)}(\gamma,\lambda_1,\lambda_2)=\mathrm{median}(\{\mathrm{MAR}_{m}^{(k)}(\gamma,\lambda_1,\lambda_2):m = 1, \dots, M\}). $
Here, $\mathrm{median}\{\cdot\}$ denotes the median of the given values, and $n_{m,k}$ denotes the sample size of the $k$th fold for client $m$.
Finally, the CV score for each candidate triplet $(\gamma,\lambda_1,\lambda_2)$ is defined as $\mathrm{CV}(\gamma,\lambda_1,\lambda_2)=\frac{1}{5}\sum_{k=1}^{5}\mathrm{MAR}^{(k)}(\gamma,\lambda_1,\lambda_2).$
The optimal tuning parameters are selected as the candidate triplet that minimizes this CV score over $\mathcal{T}$, namely,
$(\gamma^{*},\lambda_1^{*},\lambda_2^{*})=\arg\min_{(\gamma,\lambda_1,\lambda_2)\in\mathcal{T}}\mathrm{CV}(\gamma,\lambda_1,\lambda_2).$

\begin{algorithm}[tbp]
\caption{Selection of Tuning Parameters via $5$-fold CV}
\label{alg:cv_tuning}
\begin{algorithmic}[1]
\REQUIRE Local datasets $\mathcal{D}_1, \dots, \mathcal{D}_M$ and candidate sets $\Gamma, \Lambda_1, \Lambda_2$;
\STATE \textbf{Initialization:} Construct $\mathcal{T}=\Gamma \times \Lambda_1 \times \Lambda_2$; split each $\mathcal{D}_m$ into $5$ folds, denoted by $\mathcal{D}_m^{(1)},\dots,\mathcal{D}_m^{(5)}$;
\FOR{each $(\gamma,\lambda_1,\lambda_2)\in\mathcal{T}$}
    \STATE Set $\mathrm{CV}(\gamma,\lambda_1,\lambda_2)=0$;
    \FOR{$k = 1,2,\dots,5$}
        \STATE \textbf{Clients $m \in [M]$} do in parallel:
        \STATE \quad Set $\mathcal{D}_{m,\mathrm{test}}^{(k)}=\mathcal{D}_m^{(k)}$ and $\mathcal{D}_{m,\mathrm{train}}^{(k)}=\displaystyle\bigcup_{\ell\neq k}\mathcal{D}_m^{(\ell)}$;

        \STATE \textbf{Server and Clients:}
        \STATE \quad Fit Algorithm~\ref{alg:proposed} using $\{\mathcal{D}_{m,\mathrm{train}}^{(k)}\}_{m=1}^{M}$ with $(\gamma,\lambda_1,\lambda_2)$ to obtain $\hat{\bm{\beta}}$;

        \STATE \textbf{Clients $m \in [M]$} do in parallel:
        \STATE \quad Compute $\mathrm{MAR}_{m}^{(k)}(\gamma,\lambda_1,\lambda_2)$ on $\mathcal{D}_{m,\mathrm{test}}^{(k)}$ using $\hat{\bm{\beta}}_m$;
        \STATE \quad Send $\mathrm{MAR}_{m}^{(k)}(\gamma,\lambda_1,\lambda_2)$ to the server;

        \STATE \textbf{Server:}
        \STATE \quad Compute $\mathrm{MAR}^{(k)}(\gamma,\lambda_1,\lambda_2)$ using $\{\mathrm{MAR}_{m}^{(k)}(\gamma,\lambda_1,\lambda_2)\}_{m=1}^{M}$;
        \STATE \quad Update $\mathrm{CV}(\gamma,\lambda_1,\lambda_2)=\mathrm{CV}(\gamma,\lambda_1,\lambda_2)+\dfrac{1}{5}\mathrm{MAR}^{(k)}(\gamma,\lambda_1,\lambda_2)$;
    \ENDFOR
\ENDFOR
\STATE Set
$(\gamma^{*},\lambda_1^{*},\lambda_2^{*})
= \arg\min_{(\gamma,\lambda_1,\lambda_2)\in\mathcal{T}}
\mathrm{CV}(\gamma,\lambda_1,\lambda_2).$
\ENSURE Optimal tuning parameters $(\gamma^{*},\lambda_1^{*},\lambda_2^{*})$.
\end{algorithmic}
\end{algorithm}

\section{Numerical Simulation}
In this section, we evaluate the effectiveness of $\gamma$-PFL through synthetic data experiments that account for statistical heterogeneity across clients and within-client contamination. Section~4.1 examines the effectiveness of PFL in accounting for similarities across clients under statistical heterogeneity. Section~4.2 investigates the robustness of $\gamma$-PFL to within-client contamination and examines how contamination conditions and data scale affect estimation accuracy under contamination.

\paragraph{Data Generation:}
For each client $m$, let $\bm{\varepsilon}_m\in\mathbb{R}^{n_m}$ denote the error vector, $\bm{y}_m\in\mathbb{R}^{n_m}$ denote the response vector, and $\bm{X}_m\in\mathbb{R}^{n_m\times p}$ denote the covariate matrix. The data are generated from the linear model $\bm{y}_m=\bm{X}_m\bm{\beta}_m^*+\bm{\varepsilon}_m,\
m \in [M]$. Here, $\bm{\beta}_m^{*} \in \mathbb{R}^{p}$ denotes the true regression coefficient vector. To model statistical heterogeneity across clients, we set $\bm{\beta}_m^{*} = (-5, 10, \bm{0}_{p - 2})^\top$ for clients $\{1, \dots, M/2\}$ and $\bm{\beta}_m^{*} = (5, 10, \bm{0}_{p - 2})^\top$ for clients $\{M/2 + 1, \dots, M\}$. $\bm{X}_m$ is generated from a multivariate normal distribution with mean $\bm{0}_p$ and covariance structure $\mathrm{cov}(x_{is}, x_{it}) = 0.3$ for $s \ne t$ and $\mathrm{cov}(x_{is}, x_{is}) = 1$ for $s, t \in [p]$.

\paragraph{Performance Measurements:}
We use four performance measurements: root mean squared error (RMSE), recall, precision, and the $F_1$-score. The performance measures used in each experiment are selected according to the objective of the analysis. 
RMSE is used to evaluate the estimation accuracy of the true regression coefficients, whereas recall, precision, and the $F_1$-score are used to evaluate sparsity recovery.
First, we define RMSE. Let $\hat{\bm{\beta}}_{m}=(\hat{\beta}_{m1}, \dots, \hat{\beta}_{mp})^\top$ denote the estimator for client $m$, and let $\bm{\beta}_{m}^{*}=(\beta_{m1}^{*}, \dots, \beta_{mp}^{*})^\top$ denote the corresponding true value.
The RMSE is defined as $\mathrm{RMSE}=\{\frac{1}{Mp}\sum_{m=1}^{M}\sum_{j=1}^{p}(\hat{\beta}_{mj}-\beta_{mj}^{*})^{2}\}^{1/2}$.
RMSE takes nonnegative real values, and smaller values indicate more accurate recovery of the true regression coefficients.
Next, we define recall, precision, and the $F_1$-score. For client $m$, let $\mathrm{Recall}_m$ denote the fraction of truly nonzero coefficients that are estimated as nonzero, and let $\mathrm{Precision}_m$ denote the fraction of estimated nonzero coefficients that are truly nonzero.
The $F_1$-score for client $m$ is defined as
$F_{1,m}=2\left(\mathrm{Recall}_{m}^{-1}+\mathrm{Precision}_{m}^{-1}\right)^{-1}$.
The overall recall, precision, and $F_1$-score are obtained by averaging $\mathrm{Recall}_m$, $\mathrm{Precision}_m$, and $F_{1,m}$ over all clients, respectively.
These three measures range from $0$ to $1$, and values closer to $1$ indicate better sparsity recovery.

\subsection{Effectiveness of PFL}
In this experiment, we examine the effectiveness of $\gamma$-PFL, which accounts for similarities across clients under statistical heterogeneity.
We use two comparison methods adapted to the $\gamma$-divergence-based estimation framework.
The first is \textit{$\gamma$-FedAvg}, obtained by applying the $\gamma$-divergence-based loss to \textit{FedAvg}~\cite{mcmahan2017communication}, which assumes a common model shared across all clients.
The second is \textit{$\gamma$-Individual}, which applies the same $\gamma$-divergence-based estimation independently to each client using only its local data.
The detailed optimization algorithms for these comparison methods are provided in the Appendix.
We consider three combinations of the number of clients and the sample size per client:
$(M,n_m)\in\{(10,50),(10,100),(20,100)\}$. The dimension is fixed at $p=15$. In the setting without contamination, the error term follows $\varepsilon_{mi}\sim\mathcal{N}(0,1)$, whereas in the 20\% contamination setting, it follows $\varepsilon_{mi}\sim0.8\mathcal{N}(0,1)+0.2\mathcal{N}(10,0.25)$.
Table~\ref{tab:simulation_4_1_summary} summarizes the median results over 50 replications for the three methods with the Lasso and SCAD regularizers. Hereafter, L1 denotes the Lasso regularizer.

\begin{table}[tbp]
\centering
\caption{Performance comparison without and with 20\% contamination.}
\label{tab:simulation_4_1_summary}
\small
\begin{tabular}{llccccccccc}
\toprule
& & & \multicolumn{4}{c}{Without contamination}
& \multicolumn{4}{c}{With 20\% contamination} \\
\cmidrule(lr){4-7} \cmidrule(lr){8-11}
Method & Regularizer & ($M,n_m$)
& RMSE & Recall & Precision & $F_1$
& RMSE & Recall & Precision & $F_1$ \\
\midrule
\textit{$\gamma$-Individual}   & L1   & (10,50)
& 0.209 & 1.000 & 0.268 & 0.414
& 1.130 & 0.850 & 0.307 & 0.367 \\

\textit{$\gamma$-Individual}   & SCAD & (10,50)
& 0.169 & 1.000 & 0.717 & 0.807
& 1.016 & 0.850 & 0.410 & 0.413 \\

\textit{$\gamma$-FedAvg}       & L1   & (10,50)
& 1.657 & 0.500 & 1.000 & 0.667
& 1.581 & 1.000 & 0.167 & 0.286 \\

\textit{$\gamma$-FedAvg}       & SCAD & (10,50)
& 1.521 & 0.500 & 1.000 & 0.667
& 1.545 & 0.500 & 1.000 & 0.667 \\

$\gamma$-PFL & L1   & (10,50)
& 0.103 & 1.000 & 0.148 & 0.258
& 0.272 & 1.000 & 0.135 & 0.238 \\

$\gamma$-PFL & SCAD & (10,50)
& 0.024 & 1.000 & 1.000 & 1.000
& 0.037 & 1.000 & 1.000 & 1.000 \\

\midrule
\textit{$\gamma$-Individual}   & L1   & (10,100)
& 0.145 & 1.000 & 0.408 & 0.569
& 0.245 & 1.000 & 0.163 & 0.280 \\

\textit{$\gamma$-Individual}   & SCAD & (10,100)
& 0.108 & 1.000 & 0.720 & 0.820
& 0.301 & 1.000 & 0.178 & 0.301 \\

\textit{$\gamma$-FedAvg}       & L1   & (10,100)
& 1.525 & 0.500 & 0.250 & 0.333
& 1.515 & 0.500 & 0.333 & 0.400 \\

\textit{$\gamma$-FedAvg}       & SCAD & (10,100)
& 1.497 & 0.500 & 1.000 & 0.667
& 1.470 & 0.500 & 1.000 & 0.667 \\

$\gamma$-PFL & L1   & (10,100)
& 0.065 & 1.000 & 0.380 & 0.544
& 0.090 & 1.000 & 0.154 & 0.267 \\

$\gamma$-PFL & SCAD & (10,100)
& 0.015 & 1.000 & 1.000 & 1.000
& 0.022 & 1.000 & 1.000 & 1.000 \\

\midrule
\textit{$\gamma$-Individual}   & L1   & (20,100)
& 0.154 & 1.000 & 0.205 & 0.339
& 0.430 & 1.000 & 0.216 & 0.341 \\

\textit{$\gamma$-Individual}   & SCAD & (20,100)
& 0.096 & 1.000 & 0.825 & 0.890
& 0.303 & 1.000 & 0.179 & 0.302 \\

\textit{$\gamma$-FedAvg}       & L1   & (20,100)
& 1.534 & 0.500 & 0.500 & 0.500
& 1.515 & 0.500 & 0.500 & 0.500 \\

\textit{$\gamma$-FedAvg}       & SCAD & (20,100)
& 1.495 & 0.500 & 1.000 & 0.667
& 1.467 & 0.500 & 1.000 & 0.667 \\

$\gamma$-PFL & L1   & (20,100)
& 0.100 & 1.000 & 0.191 & 0.319
& 0.149 & 1.000 & 0.164 & 0.281 \\

$\gamma$-PFL & SCAD & (20,100)
& 0.014 & 1.000 & 1.000 & 1.000
& 0.126 & 1.000 & 0.250 & 0.400 \\
\bottomrule
\end{tabular}
\end{table}

As shown in Table~\ref{tab:simulation_4_1_summary}, when methods employing the same regularizer are compared, $\gamma$-PFL achieves the lowest RMSE across all experimental settings. This improvement may be attributed to its ability to account for similarities among clients and leverage shared information through joint estimation. In contrast, \textit{$\gamma$-Individual} estimates a separate model for each client, thereby limiting the amount of data available for estimating each client-specific model and resulting in lower estimation accuracy than that of $\gamma$-PFL. \textit{$\gamma$-FedAvg}, on the other hand, assumes a single global model shared across all clients. Moreover, in many settings, the RMSE tends to decrease as the sample size at each client increases. When the sample size is held fixed, however, the RMSE tends to increase with the number of clients in some settings. With respect to the effect of contamination, a comparison between the settings without contamination and those with 20\% contamination shows that the RMSE generally increases under contamination. In particular, \textit{$\gamma$-Individual} exhibits a substantial increase in RMSE, suggesting that it is relatively susceptible to the influence of outliers. Although the RMSE of \textit{$\gamma$-FedAvg} changes only slightly under contamination, it remains relatively large even in the absence of contamination. By comparison, $\gamma$-PFL exhibits a smaller increase in RMSE under contamination than \textit{$\gamma$-Individual}. These results suggest that federated learning methods that leverage information across clients tend to exhibit a smaller increase in RMSE under contamination than \textit{$\gamma$-Individual}. Regarding the $F_1$-score, $\gamma$-PFL with the SCAD regularizer achieves the highest value in most settings. In particular, in all settings without contamination, the $F_1$-score is 1.000, indicating exact recovery of the underlying sparse structure.

\subsection{Estimation Accuracy under Within-Client Contamination}
In this subsection, we evaluate the estimation accuracy of $\gamma$-PFL under within-client contamination. For comparison, we consider PerFL-RSR~\cite{liu2025robust}, a robust PFL method that employs the Huber loss~\cite{huber1964robust} to accommodate heavy-tailed errors and outliers. We first investigate the effects of the outlier magnitude and contamination proportion. We subsequently fix the contamination conditions and examine the effects of data scale by varying the number of clients, the sample size per client, and the dimension. In both experiments, estimation accuracy is evaluated using the RMSE.

\subsubsection{Effects of Contamination Conditions}
\begin{figure}[htbp]
  \centering
  \begin{subfigure}{0.32\linewidth}
    \centering
    \includegraphics[width=\linewidth]{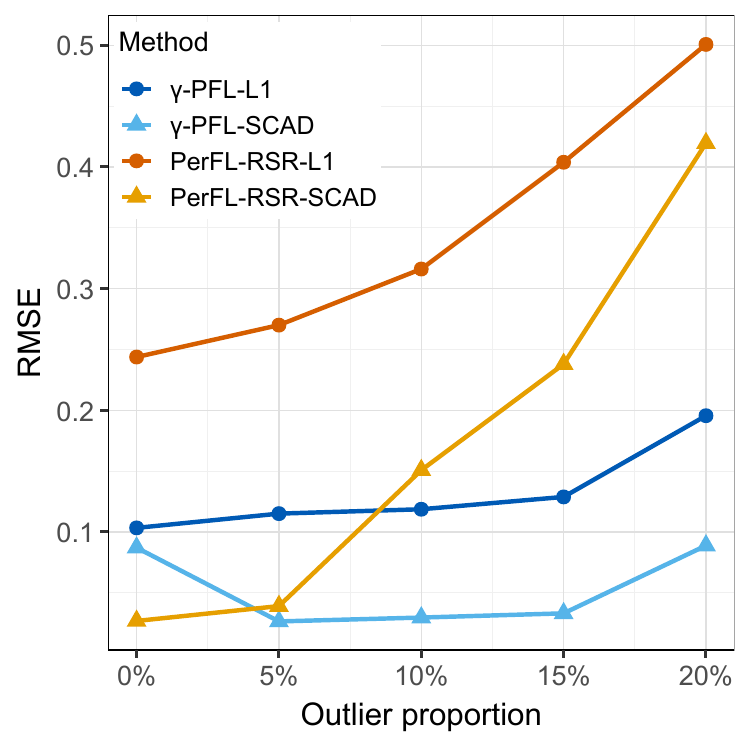}
    \caption{$\mu=10$}
    \label{fig:rmse_outlier_mu10}
  \end{subfigure}
  \hfill
  \begin{subfigure}{0.32\linewidth}
    \centering
    \includegraphics[width=\linewidth]{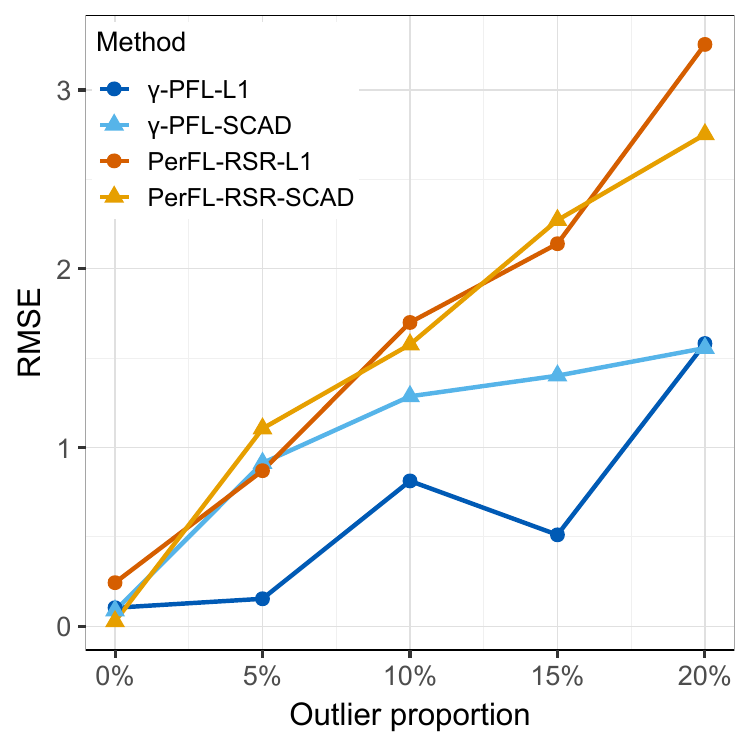}
    \caption{$\mu=100$}
    \label{fig:rmse_outlier_mu100}
  \end{subfigure}
  \hfill
  \begin{subfigure}{0.32\linewidth}
    \centering
    \includegraphics[width=\linewidth]{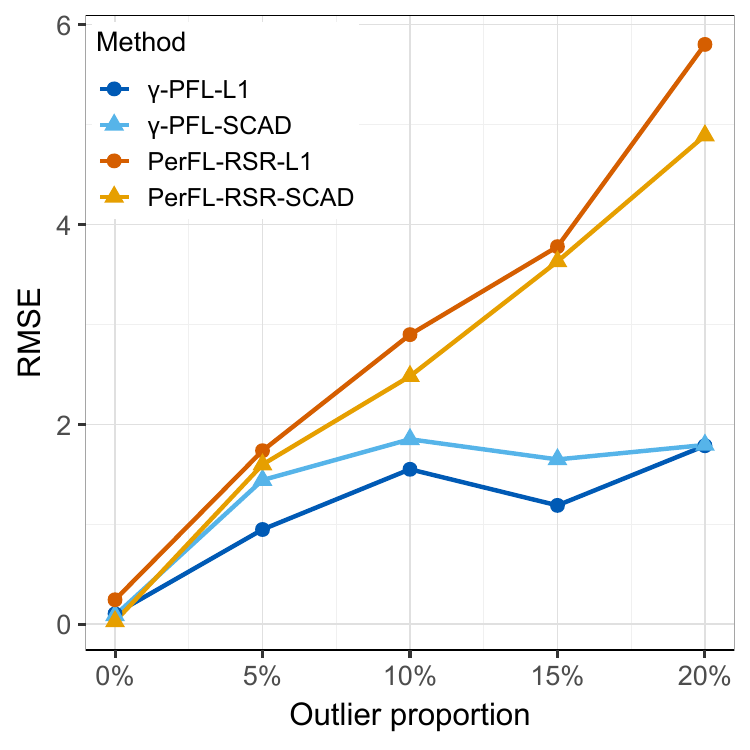}
    \caption{$\mu=200$}
    \label{fig:rmse_outlier_mu200}
  \end{subfigure}
  \caption{RMSE under different outlier magnitudes.}
  \label{fig:rmse_contamination}
\end{figure}

In this experiment, we investigate how the outlier magnitude and contamination proportion affect the estimation accuracy of $\gamma$-PFL. We fix the number of clients, the sample size per client, and the dimension at $M=10$, $n_m=50$, and $p=15$, respectively. To model within-client contamination by an outlier distribution, we assume that the error term follows $\varepsilon_{mi}\sim(1-\epsilon)\mathcal{N}(0,1)+\epsilon\mathcal{N}(\mu,0.25)$. The outlier magnitude is set to $\mu\in\{10,100,200\}$, and the contamination proportion is set to $\epsilon\in\{0.05,0.1,0.2\}$. For each combination of $\mu$ and $\epsilon$, the simulation is repeated 50 times, and the median RMSE over the 50 replications is presented in Figure~\ref{fig:rmse_contamination}.

As shown in Figure~\ref{fig:rmse_contamination}, the RMSE generally increases with either the contamination proportion $\epsilon$ or the outlier magnitude $\mu$. However, $\gamma$-PFL shows a smaller increase in RMSE than PerFL-RSR and maintains relatively stable estimation accuracy as the degree of contamination increases.

\subsubsection{Effects of Data Scale}
\begin{figure}[htbp]
\centering
\begin{subfigure}{0.31\linewidth}
\centering
\includegraphics[width=\linewidth]
{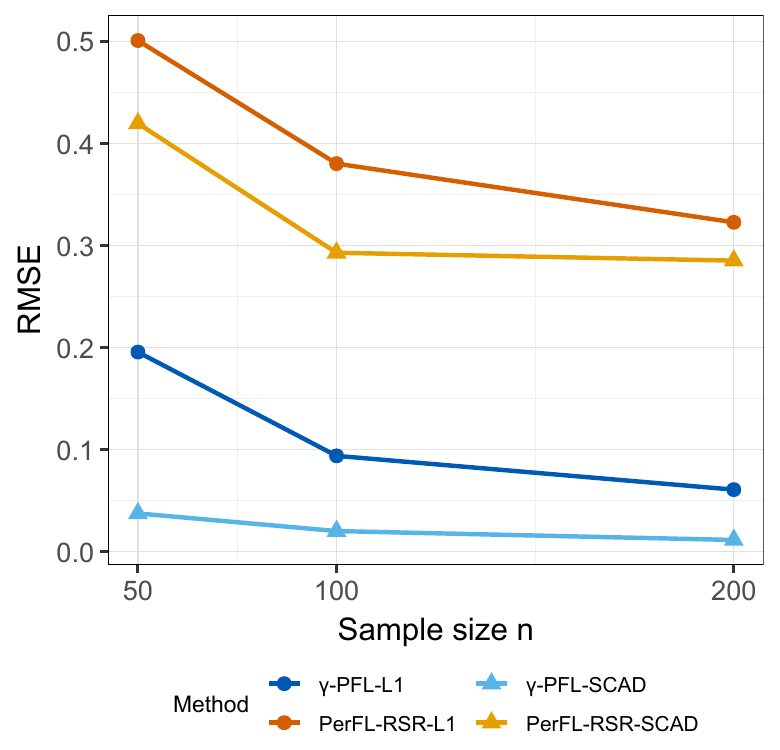}
\caption{$n_m$ varied}
\label{fig:rmse_vary_sample_size}
\end{subfigure}
\hfill
\begin{subfigure}{0.31\linewidth}
\centering
\includegraphics[width=\linewidth]
{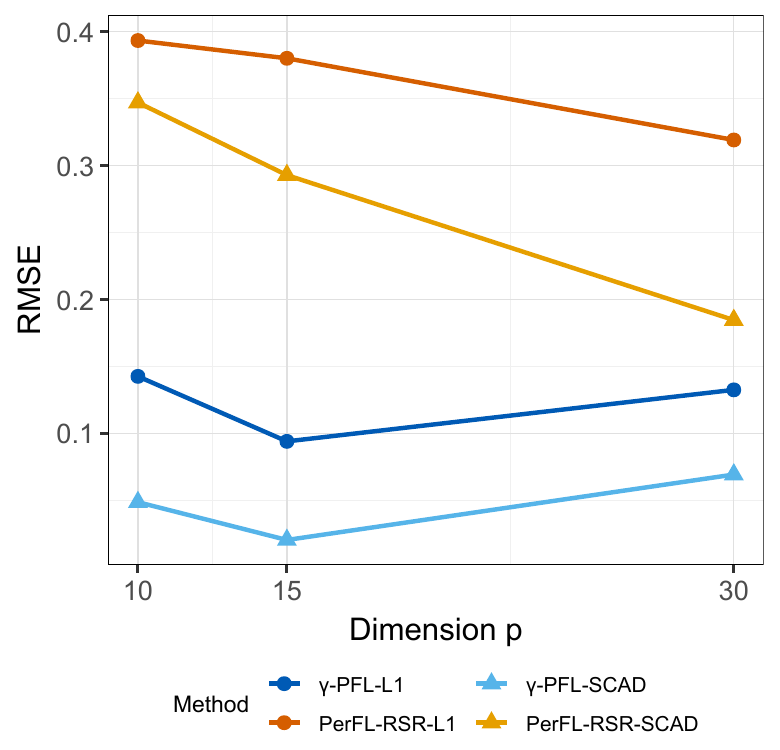}
\caption{$p$ varied ($n_m=100$)}
\label{fig:rmse_vary_dimension_n100}
\end{subfigure}
\hfill
\begin{subfigure}{0.31\linewidth}
\centering
\includegraphics[width=\linewidth]
{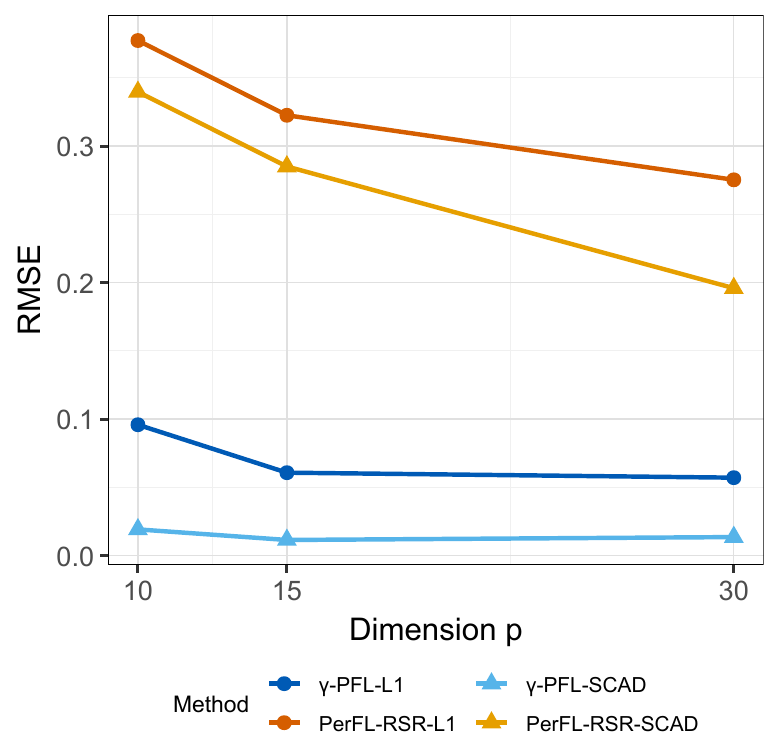}
\caption{$p$ varied ($n_m=200$)}
\label{fig:rmse_vary_dimension_n200}
\end{subfigure}
\vspace{0.5em}
\begin{subfigure}{0.31\linewidth}
\centering
\includegraphics[width=\linewidth]
{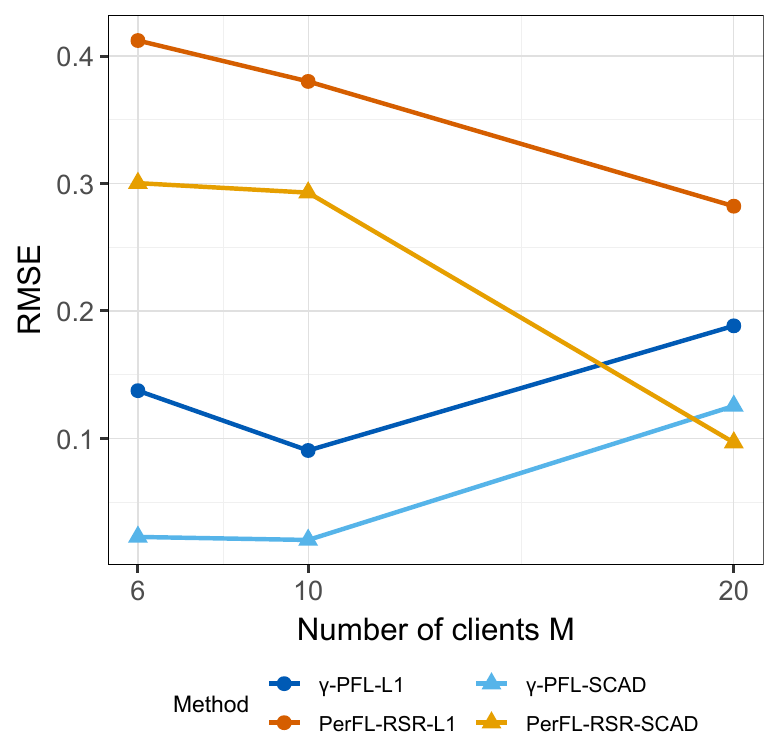}
\caption{$M$ varied ($n_m=100$)}
\label{fig:rmse_vary_clients_n100}
\end{subfigure}
\hspace{0.04\linewidth}
\begin{subfigure}{0.31\linewidth}
\centering
\includegraphics[width=\linewidth]
{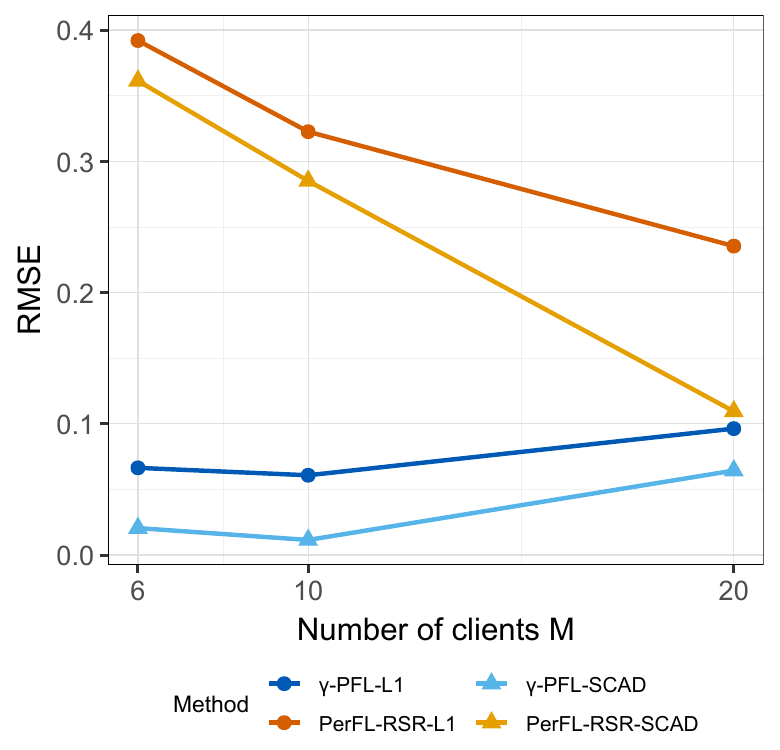}
\caption{$M$ varied ($n_m=200$)}
\label{fig:rmse_vary_clients_n200}
\end{subfigure}
\caption{RMSE under different simulation settings.}
\label{fig:rmse_vary_settings}
\end{figure}

In this experiment, we investigate how changes in data scale affect the estimation accuracy of $\gamma$-PFL under within-client contamination. To fix the contamination conditions, we assume that the error term follows
$\varepsilon_{mi}\sim 0.8\mathcal{N}(0,1)+0.2\mathcal{N}(10,0.25)$.
Under this setting, we examine the effects of the sample size per client $n_m$, the dimension $p$, and the number of clients $M$ on estimation accuracy. For each setting, the simulation is repeated 50 times, and the performance is evaluated using the median RMSE over the 50 replications.
To examine the effect of the sample size per client, we set $n_m\in\{50,100,200\}$ while fixing $M=10$ and $p=15$. The results are shown in Figure~\ref{fig:rmse_vary_settings}(a).
To examine the effect of the dimension, we set $p\in\{10,15,30\}$ for each of $n_m=100$ and $n_m=200$, while fixing $M=10$. The results for $n_m=100$ and $n_m=200$ are shown in Figure~\ref{fig:rmse_vary_settings}(b) and (c), respectively.
Finally, to examine the effect of the number of clients, we set $M\in\{6,10,20\}$ for each of $n_m=100$ and $n_m=200$, while fixing $p=15$. The results for $n_m=100$ and $n_m=200$ are shown in Figure~\ref{fig:rmse_vary_settings}(d) and (e), respectively.

As shown in Figure~\ref{fig:rmse_vary_settings}(a), the RMSE of $\gamma$-PFL decreases as the sample size per client $n_m$ increases. In addition, $\gamma$-PFL achieves lower RMSE than PerFL-RSR for all sample sizes considered.
Next, we examine the effect of the dimension $p$ by comparing Figure~\ref{fig:rmse_vary_settings}(b) and (c). When $n_m=100$, the RMSE of $\gamma$-PFL decreases as $p$ increases from 10 to 15, but increases at $p=30$, indicating a deterioration in estimation accuracy in the higher-dimensional setting. In contrast, when $n_m=200$, no clear increase in RMSE is observed as $p$ increases. Moreover, for both $n_m=100$ and $n_m=200$, $\gamma$-PFL achieves lower RMSE than PerFL-RSR across the considered dimensions.
Finally, regarding the effect of the number of clients $M$, Figure~\ref{fig:rmse_vary_settings}(d) and (e) show that, for $\gamma$-PFL, the RMSE at $M=20$ is higher than that at $M=10$ for both $n_m=100$ and $n_m=200$. Although $\gamma$-PFL achieves lower RMSE than PerFL-RSR in most settings, PerFL-RSR with the SCAD regularizer achieves a lower RMSE than $\gamma$-PFL when $n_m=100$ and $M=20$.

\section{Real Data Analysis}
In this section, we evaluate the effectiveness of $\gamma$-PFL through a real data analysis. We use the Communities and Crime dataset from the UCI Machine Learning Repository~\cite{redmond2009}. This dataset consists of data collected from communities across the United States and contains crime-related information as well as various socioeconomic variables, including population composition, median household income, and unemployment rate. In this analysis, we apply the same preprocessing procedure as Liu et al.~\cite{liu2025robust} and evaluate the predictive performance of $\gamma$-PFL.
We use \texttt{ViolentCrimesPerPop}, which represents the number of violent crimes per population, as the response variable and construct a linear regression model with 99 covariates for each client. Thus, including the intercept, 100 regression coefficients are estimated for each client. We define each combination of county and state as a client and treat each community as a sample belonging to the corresponding client.
Since some of the tuning parameters of $\gamma$-PFL are selected using 5-fold CV, a sufficient sample size is required for each client. Therefore, we restrict the analysis to the 10 clients with the largest sample sizes. Summary statistics of the sample sizes for the 10 selected clients are presented in Table~\ref{tab:top10_client_sample_size_summary}.
\begin{table}[H]
\centering
\caption{Summary statistics of the sample sizes for the 10 selected clients.}
\label{tab:top10_client_sample_size_summary}
\small
\begin{tabular}{cccccc}
\toprule
Minimum & First quartile & Median & Mean & Third quartile & Maximum \\
\midrule
18 & 18.25 & 19.5 & 27.8 & 30.5 & 76 \\
\bottomrule
\end{tabular}
\end{table}

For comparison, we use PerFL-RSR~\cite{liu2025robust} and evaluate the predictive performance of both $\gamma$-PFL and PerFL-RSR with the Lasso and SCAD regularizers.
We randomly split the data within each client into training and test sets,
using 80\% of the observations for training and the remaining 20\% for testing.
To evaluate the predictive performance, we use the root mean squared error
(RMSE).
Let $n_m^{\mathrm{test}}$ denote the number of observations in the test set
for client $m$, and let $y_{mi}$ and $\hat{y}_{mi}$ denote the observed and
predicted values, respectively, for the $i$th observation in the test set.
The RMSE for client $m$ is defined as
$\mathrm{RMSE}_{m}=\sqrt{\frac{1}{n_m^{\mathrm{test}}}\sum_{i=1}^{n_m^{\mathrm{test}}}\left(\hat{y}_{mi}-y_{mi}\right)^2}.$
For each replication, we calculate the median of $\mathrm{RMSE}_{m}$ across
all clients.
The data splitting and model estimation procedures are repeated 10 times,
and the median RMSE obtained across clients in each replication is further
aggregated by taking the median over the 10 replications.
The results are presented in Table~\ref{tab:real_data_top10_summary}.

\begin{table}[H]
\centering
\caption{Summary of RMSE for the top 10 clients based on median aggregation.}
\label{tab:real_data_top10_summary}
\begin{tabular}{llc}
\toprule
Method & Regularizer & RMSE  \\
\midrule
PerFL-RSR & L1 & 0.354  \\
PerFL-RSR & SCAD  & 0.378  \\
$\gamma$-PFL & L1 & 0.348 \\
$\gamma$-PFL & SCAD  & 0.302  \\
\bottomrule
\end{tabular}
\end{table}

As shown in Table~\ref{tab:real_data_top10_summary}, when the Lasso regularizer is used, the RMSEs of $\gamma$-PFL and PerFL-RSR are 0.348 and 0.354, respectively, indicating comparable predictive performance between the two methods.
In contrast, when the SCAD regularizer is used, $\gamma$-PFL achieves an RMSE of 0.302, which is lower than the RMSE of 0.378 obtained by PerFL-RSR.
Moreover, for $\gamma$-PFL, the RMSE with the SCAD regularizer is lower than that with the Lasso regularizer, yielding the lowest RMSE among all the methods and regularizers considered.
These results suggest that $\gamma$-PFL achieves predictive performance comparable to or better than that of PerFL-RSR in this real data analysis, particularly when combined with the SCAD regularizer.

\section{Conclusion and Discussion}
In this study, we proposed a PFL method, termed $\gamma$-PFL, that simultaneously accounts for statistical heterogeneity across clients, within-client data contamination, and model sparsity.
Specifically, we constructed an objective function that combines a loss function based on the $\gamma$-divergence to achieve robustness against within-client data contamination, a fusion regularization term to account for similarities in regression coefficients across clients, and a sparse regularization term for sparsity recovery.
Furthermore, to efficiently optimize the proposed objective function in the federated server--client system, we developed an optimization algorithm for $\gamma$-PFL that combines ADMM with the MM algorithm. 
In the numerical simulations, we confirmed that exploiting similarities across clients resulted in higher estimation accuracy than \textit{$\gamma$-Individual} and \textit{$\gamma$-FedAvg}.
In addition, in the experiments with varying within-client contamination conditions, $\gamma$-PFL exhibited smaller increases in RMSE and maintained higher estimation accuracy than the Huber-loss-based PerFL-RSR as the contamination proportion and the magnitude of outliers increased.
These results indicate that $\gamma$-PFL is particularly effective in settings where within-client data contain a high proportion of contamination or large outliers.
Furthermore, in the experiments that varied the data scale under within-client contamination, $\gamma$-PFL achieved higher estimation accuracy than PerFL-RSR in most settings with varying sample sizes per client, dimensions, and numbers of clients.
In the real data analysis, $\gamma$-PFL also achieved RMSE values comparable to or lower than those of PerFL-RSR; in particular, with the SCAD regularizer, $\gamma$-PFL achieved the lowest RMSE among the methods considered.
Overall, the numerical simulations demonstrated the effectiveness of $\gamma$-PFL as a robust and sparse PFL method that accounts for both statistical heterogeneity across clients and within-client data contamination, while the real data analysis further demonstrated its potential applicability to real-world data.

\begin{itemize}
\item The objective function of $\gamma$-PFL includes a nonconvex loss function based on the $\gamma$-divergence, and the regularization terms are also nonconvex when the SCAD regularizer is used. For PerFL-RSR, it has been shown that, under certain conditions, a subsequence of the generated parameter sequence converges to a stationary point satisfying the KKT conditions, even in the presence of nonconvex regularization terms \cite{liu2025robust}.  Therefore, establishing convergence of the proposed optimization algorithm to a KKT stationary point of the proposed nonconvex optimization problem is left for future work.

\item In the numerical simulations, the RMSE of $\gamma$-PFL increased in some settings as the number of clients increased. These results suggest that the effects of increasing the number of clients on estimation accuracy and the underlying causes require further investigation. In addition, this study assumes that all clients participate in every communication round, although this may be difficult in practice owing to constraints on network connectivity and computational resources. PerFL-RSR accommodates partial participation by randomly selecting clients in each communication round and retaining the regression coefficients of nonparticipating clients at their previous values \cite{liu2025robust}. Therefore, extending $\gamma$-PFL to accommodate partial participation and investigating the effects of the number of clients and the participation rate on estimation accuracy and computational and communication efficiency are left for future work.

\item In the tuning procedure for $\gamma$-PFL, the regularization parameters $\lambda_1$ and $\lambda_2$, as well as the robustness parameter $\gamma$, are jointly selected by 5-fold CV. This procedure incurs computational costs because model estimation must be repeated for each candidate combination. In addition, when the sample size at each client is small, dividing the local dataset into folds makes it difficult to ensure a sufficient number of observations in each fold. PerFL-RSR uses the Bayesian information criterion (BIC) to select $\lambda_1$ and $\lambda_2$ \cite{liu2025robust}. A selection criterion based on the Hyv\"arinen score has also been proposed for $\gamma$ \cite{sugasawa2021selection}. Since these selection criteria do not require data splitting, they may reduce the computational costs associated with CV and alleviate the sample size constraints imposed by data splitting. Therefore, investigating the selection of $\lambda_1$ and $\lambda_2$ using a BIC-type information criterion suitable for $\gamma$-PFL and extending the selection criterion for $\gamma$ based on the Hyv\"arinen score to the PFL setting are left for future work.
\end{itemize}
\bibliographystyle{unsrt}
\bibliography{references}
\newpage

\section*{Appendix}
\section*{A\ \ Proof of \lemref{lem:majorizing_function}}
\begin{proof}
The function $f_{\gamma}(\bm{\beta})$ can be expressed using the client-specific loss function
$L_m(\bm{\theta}_m)$ defined in \eqref{eq:client_loss} for each client $m$ as follows:
\begin{align}
  f_{\gamma}(\bm{\beta})=\frac{1}{M}\sum_{m=1}^{M}L_m(\bm{\theta}_m).
  \label{eq:lemma1_client_average}
\end{align}

For the first term of $L_m(\bm{\theta}_m)$, introducing arbitrary weights
$w_{mi}\ge0$ satisfying $\sum_{i=1}^{n_m}w_{mi}=1$ yields the following relation:
\begin{align}
  -\frac{1}{\gamma}\log\!\Bigg\{\frac{1}{n_m}\sum_{i=1}^{n_m}
  f_{y\mid x}\!\bigl(y_{mi}\mid\bm{x}_{mi};\,\bm{\theta}_m\bigr)^\gamma\Bigg\}
  &= -\frac{1}{\gamma}\log\!\Bigg\{\sum_{i=1}^{n_m}w_{mi}
  \frac{f_{y\mid x}\!\bigl(y_{mi}\mid\bm{x}_{mi};\,\bm{\theta}_m\bigr)^\gamma}{w_{mi}}\Bigg\}
  +\frac{1}{\gamma}\log n_m
  \label{eq:lemma1_jensen_identity}\\
  &\le -\frac{1}{\gamma} \sum_{i=1}^{n_m} w_{mi} \log\!\Bigg\{
  \frac{f_{y\mid x}\!\bigl( y_{mi}\mid\bm{x}_{mi};\,\bm{\theta}_m \bigr)^\gamma}{w_{mi}}
  \Bigg\}
  + \frac{1}{\gamma}\log n_m
  \label{eq:lemma1_jensen_bound}\\
  &= -\sum_{i=1}^{n_m} w_{mi} \log f_{y\mid x}\!\bigl( y_{mi}\mid\bm{x}_{mi};\,\bm{\theta}_m \bigr)
  + \frac{1}{\gamma} \sum_{i=1}^{n_m}w_{mi}\log w_{mi}\notag \\
  &\qquad+ \frac{1}{\gamma}\log n_m .
  \label{eq:lemma1_jensen_expanded}
\end{align}
Here, the transformation from \eqref{eq:lemma1_jensen_identity} to
\eqref{eq:lemma1_jensen_bound} follows from Jensen's inequality.
However, to guarantee the non-increasing property of the original objective function under the MM algorithm,
the surrogate function must satisfy the touching condition, under which it coincides with the original objective function at the current iterate.
Therefore, at iteration $t$, the weight $w_{mi}^{(t)}$ satisfying
$w_{mi}^{(t)} \propto
f_{y\mid x}\!\bigl(y_{mi}\mid\bm{x}_{mi};\,\bm{\theta}_m^{(t)}\bigr)^\gamma$
as the equality condition of Jensen's inequality is determined from
$\sum_{i=1}^{n_m}w_{mi}^{(t)}=1$ as follows:
\begin{align}
  w_{mi}^{(t)} = \frac{ f_{y\mid x}\!\bigl( y_{mi}\mid \bm{x}_{mi};\,\bm{\theta}_m^{(t)} \bigr)^{\gamma} }{
  \sum_{k=1}^{n_m} f_{y\mid x}\!\bigl( y_{mk}\mid \bm{x}_{mk};\,\bm{\theta}_m^{(t)} \bigr)^{\gamma} }.\notag
\end{align}
Furthermore, the second and third terms on the right-hand side of
\eqref{eq:lemma1_jensen_expanded} are constant terms independent of
$\bm{\theta}_m$.
Therefore, by omitting these constant terms and combining the remaining term with the second term of
$L_m(\bm{\theta}_m)$,
the surrogate function
$F_{\gamma(m)}(\bm{\beta}_m,\sigma_m^2)$
for $L_m(\bm{\theta}_m)$ based on the MM algorithm is given as follows:
\begin{align}
  F_{\gamma(m)}(\bm{\beta}_m,\sigma_m^2)
  &= -\sum_{i=1}^{n_m} w_{mi}^{(t)}
  \log f_{y\mid x}\!\bigl( y_{mi}\mid\bm{x}_{mi};\,\bm{\theta}_m \bigr)
  + \frac{1}{1+\gamma} \log\!\Bigg\{ \frac{1}{n_m}\sum_{i=1}^{n_m} \int
  f_{y\mid x}\!\bigl(y\mid \bm{x}_{mi};\,\bm{\theta}_m \bigr)^{1+\gamma} \,dy \Bigg\}
  \label{eq:lemma1_surrogate_initial}\\
  &= \frac{1}{2}\log(2\pi\sigma_m^2) + \frac{1}{2\sigma_m^2} \sum_{i=1}^{n_m} w_{mi}^{(t)}
  \bigl(y_{mi}-\bm{x}_{mi}^\top\bm{\beta}_{m}\bigr)^2
  + \frac{1}{1+\gamma} \log\!\left\{ (2\pi\sigma_m^2)^{-\gamma/2} (1+\gamma)^{-1/2} \right\}
  \label{eq:lemma1_surrogate_gaussian}\\
  &= \frac{1}{2(1+\gamma)}\log\sigma_m^2 + \frac{1}{2\sigma_m^2} \sum_{i=1}^{n_m} w_{mi}^{(t)}
  \bigl(y_{mi}-\bm{x}_{mi}^\top\bm{\beta}_{m}\bigr)^2
  + \frac{1}{2(1+\gamma)} \log\left(\frac{2\pi}{1+\gamma}\right).
  \label{eq:lemma1_surrogate_local}
\end{align}
Here, in the transformation from \eqref{eq:lemma1_surrogate_initial} to
\eqref{eq:lemma1_surrogate_gaussian}, we used
$\sum_{i=1}^{n_m}w_{mi}^{(t)}=1$ and the probability density function obtained by assuming that
$f_{y\mid x}(y\mid\bm{x}_{mi};\bm{\theta}_m)$ follows the normal distribution
$\mathcal{N}(\bm{x}_{mi}^\top\bm{\beta}_m,\sigma_m^2)$.
Furthermore, in the transformation from \eqref{eq:lemma1_surrogate_gaussian} to
\eqref{eq:lemma1_surrogate_local}, the terms were rearranged using the properties of logarithms.
Accordingly, for each client $m$, a surrogate function based on the MM algorithm can be constructed as in
\eqref{eq:lemma1_surrogate_local}.
Since the third term in \eqref{eq:lemma1_surrogate_local} is a constant term independent of
$\bm{\beta}_m$ and $\sigma_m^2$, by omitting this term and averaging the surrogate functions across clients based on
\eqref{eq:lemma1_client_average}, the surrogate function
$F_\gamma(\bm{\beta},\bm{\sigma}^2)$
for $f_{\gamma}(\bm{\beta})$ is given as follows:
\begin{align}
  F_\gamma(\bm{\beta},\bm{\sigma}^2) &= \frac{1}{M} \sum_{m=1}^{M} F_{\gamma(m)}(\bm{\beta}_m,\sigma_m^2)
  \notag\\
  &= \frac{1}{M} \sum_{m=1}^{M} \Bigg\{ \frac{1}{2(1+\gamma)}\log\sigma_m^2
  + \frac{1}{2\sigma_m^2} \sum_{i=1}^{n_m} w_{mi}^{(t)}
  \bigl\{ y_{mi}-\bm{x}_{mi}^\top\bm{\beta}_{m} \bigr\}^2 \Bigg\}.\notag
\end{align}
\end{proof}

\section*{B\ \ Proof of \thmref{thm:rank_condition}}
Before proving \thmref{thm:rank_condition}, we first establish the following claim on the Lipschitz continuity of $\nabla F_{\gamma}$.
\begin{claim}\label{clm:lipschitz_gradient}
Lipschitz continuity of $\nabla F_{\gamma}$\\
For the function
$\nabla F_{\gamma}:\mathbb{R}^{pM}\rightarrow\mathbb{R}^{pM}$,
the following inequality holds for any
$\bm{u},\bm{v}\in\mathbb{R}^{pM}$:
\begin{align}
  \|\nabla F_\gamma(\bm{u})-\nabla F_\gamma(\bm{v})\|_2 \le L\|\bm{u}-\bm{v}\|_2,
  \label{eq:lipschitz_bound}
\end{align}
where the Lipschitz constant $L>0$ is given by
\begin{align}
  L=\frac{1}{M}\varsigma_{\max}(\tilde{\bm{X}}^\top\tilde{\bm{X}}).\notag
\end{align}
\end{claim}
\begin{proof}
For matrices, $\|\cdot\|_2$ denotes the spectral norm, and
$\varsigma_{\max}(\cdot)$ denotes the largest eigenvalue of a symmetric matrix.
Here, the left-hand side of \eqref{eq:lipschitz_bound} can be rewritten as follows:
\begin{align}
  \|\nabla F_\gamma(\bm{u})-\nabla F_\gamma(\bm{v})\|_2 &=
  \left\| \frac{1}{M}\tilde{\bm{X}}^\top (\tilde{\bm{X}}\bm{u}-\tilde{\bm{y}})
  - \frac{1}{M}\tilde{\bm{X}}^\top (\tilde{\bm{X}}\bm{v}-\tilde{\bm{y}}) \right\|_2
  \label{eq:lipschitz_step1}\\
  &= \frac{1}{M} \|\tilde{\bm{X}}^\top\tilde{\bm{X}}(\bm{u}-\bm{v})\|_2
  \label{eq:lipschitz_step2}\\
  &\le \frac{1}{M} \|\tilde{\bm{X}}^\top\tilde{\bm{X}}\|_2 \|\bm{u}-\bm{v}\|_2
  \label{eq:lipschitz_step3}\\
  &= \frac{1}{M} \varsigma_{\max} ( \tilde{\bm{X}}^\top\tilde{\bm{X}} ) \|\bm{u}-\bm{v}\|_2.
  \label{eq:lipschitz_step4}
\end{align}

Here, \eqref{eq:lipschitz_step1} represents the difference between the gradients obtained by setting
$\bm{\beta}=\bm{u}$ and $\bm{\beta}=\bm{v}$ in \eqref{eq:gradient_global}.
The transformation from \eqref{eq:lipschitz_step2} to \eqref{eq:lipschitz_step3} follows from the property of the spectral norm.
Furthermore, in the transformation from \eqref{eq:lipschitz_step3} to \eqref{eq:lipschitz_step4},
since $\tilde{\bm{X}}^\top\tilde{\bm{X}}$ is a symmetric positive semidefinite matrix,
its spectral norm is equal to its largest eigenvalue.
Therefore, from \eqref{eq:lipschitz_step4}, the Lipschitz constant $L$ satisfying
\eqref{eq:lipschitz_bound} is given by
\begin{align}
  L = \frac{1}{M} \varsigma_{\max} ( \tilde{\bm{X}}^\top\tilde{\bm{X}} ).\notag
\end{align}
\end{proof}

Using \claimref{clm:lipschitz_gradient}, we now prove \thmref{thm:rank_condition}.
\begin{proof}
Let $C$ be the line segment connecting $\bm{\beta}^{(t)}$ and $\bm{\beta}$, defined using the function
$\bm{z}:[0,1]\to\mathbb{R}^{pM}$ as follows:
\begin{align}
  C = \left\{ \bm{z}(s) = \bm{\beta}^{(t)}
  + s\bigl(\bm{\beta}-\bm{\beta}^{(t)}\bigr) \mid s\in[0,1] \right\}.\notag
\end{align}
In this case, the following equalities hold:
\begin{align}
  \int_C \nabla F_\gamma(\bm{z})^\top d\bm{z} &= \int_0^1 \nabla F_\gamma(\bm{z}(s))^\top \bm{z}'(s)\,ds
  \label{eq:line_integral_parameterized}\\
  &= \int_0^1 \frac{d}{ds}F_\gamma(\bm{z}(s))\,ds
  \label{eq:line_integral_chain_rule}\\
  &= F_\gamma(\bm{z}(1))-F_\gamma(\bm{z}(0)) \notag\\
  &= F_\gamma(\bm{\beta})-F_\gamma(\bm{\beta}^{(t)}).
  \label{eq:line_integral_difference}
\end{align}

The transformation to \eqref{eq:line_integral_parameterized} follows from the definition of the line integral.
Furthermore, the transformation from \eqref{eq:line_integral_parameterized} to
\eqref{eq:line_integral_chain_rule} follows from the chain rule applied to the composite function
$F_\gamma(\bm{z}(s))$.
Using \eqref{eq:line_integral_difference}, the following equalities hold:
\begin{align}
  F_\gamma(\bm{\beta})-F_\gamma(\bm{\beta}^{(t)}) &= \int_C \nabla F_\gamma(\bm{z})^\top d\bm{z} \notag\\
  &= \int_C \Bigl\{ \nabla F_\gamma(\bm{z}) -\nabla F_\gamma(\bm{\beta}^{(t)})
  +\nabla F_\gamma(\bm{\beta}^{(t)}) \Bigr\}^\top d\bm{z} \notag\\
  &= \int_C \nabla F_\gamma(\bm{\beta}^{(t)})^\top d\bm{z} + \int_C \bigl( \nabla F_\gamma(\bm{z})
  -\nabla F_\gamma(\bm{\beta}^{(t)}) \bigr)^\top d\bm{z} \notag\\
  &= \nabla F_\gamma(\bm{\beta}^{(t)})^\top \bigl( \bm{\beta}-\bm{\beta}^{(t)} \bigr)
  + \int_C \bigl( \nabla F_\gamma(\bm{z}) -\nabla F_\gamma(\bm{\beta}^{(t)}) \bigr)^\top d\bm{z}.
  \label{eq:decomposition_final}
\end{align}
Therefore, from \eqref{eq:decomposition_final},
\eqref{eq:majorization_bound} is equivalent to the following inequality:
\begin{align}
  \int_C \bigl( \nabla F_\gamma(\bm{z}) -\nabla F_\gamma(\bm{\beta}^{(t)}) \bigr)^\top d\bm{z}
  \le \frac{1}{2\tau} \|\bm{\beta}-\bm{\beta}^{(t)}\|_{\bm{H}}^{2}.
  \label{eq:integral_goal_bound}
\end{align}
The left-hand side of \eqref{eq:integral_goal_bound} can be bounded as follows:
\begin{align}
  \int_C \bigl( \nabla F_\gamma(\bm{z}) -\nabla F_\gamma(\bm{\beta}^{(t)}) \bigr)^\top d\bm{z}
  &= \int_0^1 \bigl( \nabla F_\gamma(\bm{z}(s))
  -\nabla F_\gamma(\bm{\beta}^{(t)}) \bigr)^\top \bigl( \bm{\beta}-\bm{\beta}^{(t)} \bigr)\,ds
  \label{eq:bound_parameterization}\\
  &\le \int_0^1 \|\nabla F_\gamma(\bm{z}(s))
  -\nabla F_\gamma(\bm{\beta}^{(t)})\|_2\, \|\bm{\beta}-\bm{\beta}^{(t)}\|_2\,ds
  \label{eq:bound_cauchy_schwarz}\\
  &\le L\int_0^1 \|\bm{z}(s)-\bm{\beta}^{(t)}\|_2\, \|\bm{\beta}-\bm{\beta}^{(t)}\|_2\,ds
  \label{eq:bound_lipschitz}\\
  &= L\int_0^1 s\,\|\bm{\beta}-\bm{\beta}^{(t)}\|_2^2\,ds \notag\\
  &= \frac{L}{2}\, \|\bm{\beta}-\bm{\beta}^{(t)}\|_2^2.
  \label{eq:bound_final}
\end{align}
The transformation from \eqref{eq:bound_parameterization} to \eqref{eq:bound_cauchy_schwarz}
follows from the Cauchy--Schwarz inequality.
Furthermore, the transformation from \eqref{eq:bound_cauchy_schwarz} to
\eqref{eq:bound_lipschitz} follows from the Lipschitz continuity of
$\nabla F_\gamma$ established in \claimref{clm:lipschitz_gradient}.
From \eqref{eq:integral_goal_bound} and \eqref{eq:bound_final}, a sufficient condition for
\eqref{eq:integral_goal_bound} to hold is given by
\begin{align}
  \frac{L}{2} \|\bm{\beta}-\bm{\beta}^{(t)}\|_2^2
  \le \frac{1}{2\tau} \|\bm{\beta}-\bm{\beta}^{(t)}\|_{\bm{H}}^2.
  \label{eq:sufficient_matrix_condition}
\end{align}
When $\bm{\beta}=\bm{\beta}^{(t)}$, \eqref{eq:sufficient_matrix_condition} holds trivially.
Therefore, it suffices to consider the case
$\bm{\beta}\neq\bm{\beta}^{(t)}$, for which rearranging
\eqref{eq:sufficient_matrix_condition} gives
\begin{align}
  \tau L &\le
  \frac{ \|\bm{\beta}-\bm{\beta}^{(t)}\|_{\bm{H}}^2 }
  { \|\bm{\beta}-\bm{\beta}^{(t)}\|_2^2 } \notag\\
  &= \frac{ (\bm{\beta}-\bm{\beta}^{(t)})^\top \bm{H} (\bm{\beta}-\bm{\beta}^{(t)}) }
  { \|\bm{\beta}-\bm{\beta}^{(t)}\|_2^2 }.
  \label{eq:rayleigh_quotient}
\end{align}
Here, $\varsigma_{\min}(\cdot)$ denotes the smallest eigenvalue of a symmetric matrix.
Since $\bm{H}$ is symmetric, the property of the Rayleigh quotient gives
\begin{align}
  \varsigma_{\min}(\bm{H}) \le
  \frac{ (\bm{\beta}-\bm{\beta}^{(t)})^\top \bm{H} (\bm{\beta}-\bm{\beta}^{(t)}) }
  { \|\bm{\beta}-\bm{\beta}^{(t)}\|_2^2 }.\notag
\end{align}
Therefore, a sufficient condition for \eqref{eq:rayleigh_quotient} to hold is
\begin{align}
  \tau L &\le \varsigma_{\min}(\bm{H}) \notag\\
  &= \varsigma_{\min} \left( r\bm{I}-\rho\tau\bm{A}^\top\bm{A} \right).
  \label{eq:rank_eigenvalue_condition}
\end{align}
Since $\bm{A}^\top\bm{A}$ is symmetric, it can be diagonalized using an orthogonal matrix
$\bm{Q}$ as follows.
Here, $\bm{\Lambda}$ is a diagonal matrix whose diagonal entries are the eigenvalues of
$\bm{A}^\top\bm{A}$:
\begin{align}
  \bm{A}^\top\bm{A} = \bm{Q}\bm{\Lambda}\bm{Q}^\top.
  \label{eq:ata_diagonalization}
\end{align}
Using \eqref{eq:ata_diagonalization}, $\bm{H}$ can be rewritten as
\begin{align}
  \bm{H} &= r\bm{I}-\rho\tau\bm{A}^\top\bm{A} \notag\\
  &= r\bm{Q}\bm{I}\bm{Q}^\top -\rho\tau\bm{Q}\bm{\Lambda}\bm{Q}^\top \notag\\
  &= \bm{Q} \left( r\bm{I}-\rho\tau\bm{\Lambda} \right) \bm{Q}^\top.
  \label{eq:h_diagonalization}
\end{align}
From \eqref{eq:h_diagonalization},
$r\bm{I}-\rho\tau\bm{\Lambda}$ is a diagonal matrix whose diagonal entries are the eigenvalues of
$\bm{H}$.
Therefore, since $\rho\tau>0$, the following equality holds:
\begin{align}
  \varsigma_{\min} \left( r\bm{I}-\rho\tau\bm{A}^\top\bm{A} \right) = r-\rho\tau\, \varsigma_{\max}
  \left( \bm{A}^\top\bm{A} \right).
  \label{eq:minimum_eigenvalue}
\end{align}
Therefore, rearranging the inequality obtained from \eqref{eq:rank_eigenvalue_condition} and \eqref{eq:minimum_eigenvalue}
with respect to $r$ and substituting the Lipschitz constant $L$ obtained in
\claimref{clm:lipschitz_gradient}, we obtain the following sufficient condition for
\eqref{eq:majorization_bound}:
\begin{align}
  r \ge \rho\tau\,\varsigma_{\max}(\bm{A}^\top\bm{A})
  + \frac{\tau}{M} \varsigma_{\max} ( \tilde{\bm{X}}^\top\tilde{\bm{X}} ).\notag
\end{align}

\end{proof}
\section*{C\ \ Proof of \lemref{lem:local_gradient}}
\begin{proof}
The gradient of the surrogate function $F_\gamma(\bm{\beta})$ with respect to
$\bm{\beta}$ can be expressed as follows:
\begin{align}
  \nabla F_\gamma(\bm{\beta}^{(t)}) &= \frac{\partial}{\partial\bm{\beta}} \left\{ \frac{1}{2M}
  \left\| \tilde{\bm{y}}^{(t)} -\tilde{\bm{X}}^{(t)}\bm{\beta} \right\|_2^2 \right\} \notag\\
  &= \frac{1}{M} \tilde{\bm{X}}^{(t)\top} \left( \tilde{\bm{X}}^{(t)}\bm{\beta}^{(t)}
  -\tilde{\bm{y}}^{(t)} \right)
  \label{eq:gradient_global}\\
  &= \frac{1}{M} \begin{pmatrix} \tilde{\bm{X}}_1^{(t)\top}\tilde{\bm{X}}_1^{(t)}\bm{\beta}_1^{(t)}
  -\tilde{\bm{X}}_1^{(t)\top}\tilde{\bm{y}}_1^{(t)} \\
  \tilde{\bm{X}}_2^{(t)\top}\tilde{\bm{X}}_2^{(t)}\bm{\beta}_2^{(t)}
  -\tilde{\bm{X}}_2^{(t)\top}\tilde{\bm{y}}_2^{(t)} \\
  \vdots \\
  \tilde{\bm{X}}_M^{(t)\top}\tilde{\bm{X}}_M^{(t)}\bm{\beta}_M^{(t)}
  -\tilde{\bm{X}}_M^{(t)\top}\tilde{\bm{y}}_M^{(t)} \end{pmatrix}
  \label{eq:gradient_block}\\
  &= \frac{1}{M} \begin{pmatrix} \frac{1}{\sigma_1^{2(t)}}
  \bm{X}_1^\top\bm{W}_1^{(t)}\bm{X}_1\bm{\beta}_1^{(t)}
  - \frac{1}{\sigma_1^{2(t)}} \bm{X}_1^\top\bm{W}_1^{(t)}\bm{y}_1 \\
  \frac{1}{\sigma_2^{2(t)}} \bm{X}_2^\top\bm{W}_2^{(t)}\bm{X}_2\bm{\beta}_2^{(t)}
  - \frac{1}{\sigma_2^{2(t)}} \bm{X}_2^\top\bm{W}_2^{(t)}\bm{y}_2 \\
  \vdots \\
  \frac{1}{\sigma_M^{2(t)}} \bm{X}_M^\top\bm{W}_M^{(t)}\bm{X}_M\bm{\beta}_M^{(t)}
  - \frac{1}{\sigma_M^{2(t)}} \bm{X}_M^\top\bm{W}_M^{(t)}\bm{y}_M \end{pmatrix}.
  \label{eq:gradient_local}
\end{align}
In the transformation from \eqref{eq:gradient_global} to \eqref{eq:gradient_block},
we used
$\tilde{\bm{X}}^{(t)}
= \mathrm{blkdiag}(\tilde{\bm{X}}_1^{(t)}, \tilde{\bm{X}}_2^{(t)},
\dots, \tilde{\bm{X}}_M^{(t)})$.
Furthermore, in the transformation from \eqref{eq:gradient_block} to \eqref{eq:gradient_local},
for each client $m$, we substituted
$\tilde{\bm{X}}_m^{(t)}
= (\bm{W}_m^{(t)})^{1/2}\bm{X}_m/\sigma_m^{(t)}$
and
$\tilde{\bm{y}}_m^{(t)}
= (\bm{W}_m^{(t)})^{1/2}\bm{y}_m/\sigma_m^{(t)}$.
Since
$\nabla F_\gamma(\bm{\beta}^{(t)})
= \frac{1}{M}\bigl(
\nabla F_{\gamma(1)}(\bm{\beta}_1^{(t)})^\top,
\nabla F_{\gamma(2)}(\bm{\beta}_2^{(t)})^\top,
\dots,
\nabla F_{\gamma(M)}(\bm{\beta}_M^{(t)})^\top
\bigr)^\top$,
the local gradient corresponding to client $m$,
$\nabla F_{\gamma(m)}(\bm{\beta}_m^{(t)})$,
is obtained from the correspondence with the $m$th block of
\eqref{eq:gradient_local} as follows:
\begin{align}
  \nabla F_{\gamma(m)}(\bm{\beta}_m^{(t)}) &= \frac{1}{\sigma_m^{2\,(t)}}
  \left( \bm{X}_m^\top\bm{W}_m^{(t)}\bm{X}_m\bm{\beta}_m^{(t)}
  - \bm{X}_m^\top\bm{W}_m^{(t)}\bm{y}_m \right).\notag
\end{align}
\end{proof}
\section*{D\ \ Details of Comparison Methods}
Unless otherwise stated, the notation and definitions are the same as those used in the main text.
\subsection*{D.1\ \ \textit{$\gamma$-Individual}}
\subsubsection*{Optimization Problem}
In \textit{$\gamma$-Individual}, each client $m \in [M]$ independently solves
the following optimization problem:
\begin{align}
  \hat{\bm{\beta}}_m = \mathop{\arg\min}_{\bm{\beta}_m \in \mathbb{R}^{p}} \;
  &\frac{1}{2(1+\gamma)}\log\sigma_m^2 + \frac{1}{2\sigma_m^2} \sum_{i=1}^{n_m} w_{mi}
  \bigl\{y_{mi}-\bm{x}_{mi}^\top\bm{\beta}_{m}\bigr\}^2 + \sum_{j=1}^{p} p_{\lambda}(|\beta_{mj}|).
  \label{eq:gamma_individual_objective}
\end{align}

The first and second terms are derived from the loss function based on the
$\gamma$-divergence using the MM algorithm, whereas the third term is a sparse
regularization term on the regression coefficients.
Therefore, \textit{$\gamma$-Individual} performs robust and sparse estimation of the
regression coefficients for each client in the presence of data contamination.

\subsubsection*{Update Rules Based on ADMM}
To solve the optimization problem in \eqref{eq:gamma_individual_objective}, we use ADMM.
To this end, we introduce an auxiliary variable
$\bm{\delta}_m \in \mathbb{R}^{p}$ and impose the constraint
$\bm{\beta}_m=\bm{\delta}_m$.
Here, we define
\begin{align*}
  f_{\gamma,m}(\bm{\beta}_m) &= \frac{1}{2(1+\gamma)}\log\sigma_m^2
  + \frac{1}{2\sigma_m^2} \sum_{i=1}^{n_m} w_{mi} \bigl\{y_{mi}-\bm{x}_{mi}^\top\bm{\beta}_{m}\bigr\}^2,
  \qquad h(\bm{\delta}_m) = \sum_{j=1}^{p} p_{\lambda}(|\delta_{mj}|).
\end{align*}
With the above definitions, the optimization problem in
\eqref{eq:gamma_individual_objective} can be reformulated as the following constrained optimization problem:
\begin{equation}
  \min \; f_{\gamma,m}(\bm{\beta}_m) + h(\bm{\delta}_m), \quad \text{s.t.} \quad \bm{\beta}_m
  = \bm{\delta}_m.\notag
\end{equation}
For this constrained optimization problem, the augmented Lagrangian function
$L_\rho(\bm{\beta}_m,\bm{\delta}_m,\bm{u}_m)$ is defined, with Lagrange multiplier
$\bm{u}_m\in\mathbb{R}^{p}$ and penalty parameter
$\rho\in\mathbb{R}_{>0}$, as
\begin{align}
  L_\rho(\bm{\beta}_m,\bm{\delta}_m,\bm{u}_m) =&\, f_{\gamma,m}(\bm{\beta}_m) + h(\bm{\delta}_m)
  + \bm{u}_m^\top(\bm{\beta}_m-\bm{\delta}_m) + \frac{\rho}{2} \|\bm{\beta}_m-\bm{\delta}_m\|_2^2.\notag
\end{align}

\paragraph{Update of $\bm{\beta}_m$:}
By differentiating
$L_\rho(\bm{\beta}_m,\bm{\delta}_m^{(t)},\bm{u}_m^{(t)})$
with respect to $\bm{\beta}_m$ and solving the corresponding stationary condition,
we obtain the following closed-form update:
\begin{align}
  \bm{\beta}_m^{(t+1)} &= \Bigl( \tfrac{1}{\sigma_m^{2\,(t)}} \bm{X}_m^\top \bm{W}_m^{(t)} \bm{X}_m
  +\rho\bm{I}_p \Bigr)^{-1} \Bigl( \tfrac{1}{\sigma_m^{2\,(t)}} \bm{X}_m^\top \bm{W}_m^{(t)} \bm{y}_m
  +\rho\bm{\delta}_m^{(t)} -\bm{u}_m^{(t)} \Bigr).
  \label{eq:gamma_individual_beta_update}
\end{align}
\paragraph{Update of $\bm{\delta}_m$:}
Minimizing
$L_\rho(\bm{\beta}_m^{(t+1)},\bm{\delta}_m,\bm{u}_m^{(t)})$
with respect to $\bm{\delta}_m$ gives
$\bm{\delta}_m^{(t+1)}
= \mathop{\arg\min}_{\bm{\delta}_m}\bigl\{
h(\bm{\delta}_m) +\allowbreak \frac{\rho}{2}
\|\bm{\delta}_m-\bm{\beta}_m^{(t+1)}-\rho^{-1}\bm{u}_m^{(t)}\|_2^2
\bigr\}$.
Therefore, using the proximal operator $T_{\lambda,\rho}$,
the following update is obtained for each $j \in \{1, \dots, p\}$:
\begin{align}
  \delta_{mj}^{(t+1)} = T_{\lambda,\rho} \left( \beta_{mj}^{(t+1)} +\rho^{-1}u_{mj}^{(t)} \right).
  \label{eq:gamma_individual_delta_update}
\end{align}
\paragraph{Update of $\bm{u}_m$:}
According to the update rule for the Lagrange multiplier in ADMM,
$\bm{u}_m$ is updated as follows:
\begin{align}
  \bm{u}_m^{(t+1)} = \bm{u}_m^{(t)} + \rho \bigl( \bm{\beta}_m^{(t+1)} - \bm{\delta}_m^{(t+1)} \bigr).
  \label{eq:gamma_individual_dual_update}
\end{align}
\paragraph{Update of $w_{mi}$:}
To ensure that the surrogate function obtained by applying the MM algorithm to the loss function based on the $\gamma$-divergence satisfies the touching condition at the current iterate, the weight $w_{mi}$ is updated as follows:
\begin{align}
  w_{mi}^{(t)} = \frac{ f_{y\mid x}(y_{mi}\mid \bm{x}_{mi};\,\bm{\theta}_m^{(t)})^{\gamma} } {
  \sum_{k=1}^{n_m} f_{y\mid x}(y_{mk}\mid \bm{x}_{mk};\,\bm{\theta}_m^{(t)})^{\gamma} }.
  \label{eq:gamma_individual_weight_update}
\end{align}
\paragraph{Update of $\sigma_m^2$:}
By differentiating the objective function with respect to $\sigma_m^2$
and solving the corresponding stationary condition, we obtain the following closed-form update:
\begin{align}
  \sigma_m^{2(t+1)} = (1+\gamma) \bigl(\bm{y}_m-\bm{X}_m\bm{\beta}_m^{(t+1)}\bigr)^\top \bm{W}_m^{(t)}
  \bigl(\bm{y}_m-\bm{X}_m\bm{\beta}_m^{(t+1)}\bigr).
  \label{eq:gamma_individual_sigma_update}
\end{align}

\subsubsection*{Optimization Algorithm}
Accordingly, the update scheme of \textit{$\gamma$-Individual} is summarized in
Algorithm~\ref{alg:gamma_individual}.
The initial values of $\bm{\beta}_m^{(0)}$ and $\sigma_m^{2(0)}$ are determined based on ridge regression in the same manner as in the proposed method.
\begin{algorithm}[H]
\caption{\textit{$\gamma$-Individual} Algorithm}
\label{alg:gamma_individual}
\begin{algorithmic}[1]
\REQUIRE Tuning parameters $\rho, \gamma, \lambda$,
and $\iota$ only when SCAD regularization is used; number of iterations $T$;
\STATE \textbf{Initialization:}
Set $\bm{\delta}_m^{(0)}=\bm{0}$ and $\bm{u}_m^{(0)}=\bm{0}$;
Compute the initial $\bm{\beta}_m^{(0)}$ and $\sigma_m^{2(0)}$
by ridge regression for each client $m$;
\FOR{$t = 0, 1, \dots, T - 1$}
  \STATE \textbf{Clients $m \in [M]$ do in parallel:}
  \STATE \quad Update $w_{mi}^{(t)}$ for $i = 1, \dots, n_m$ by \textup{(\ref{eq:gamma_individual_weight_update})};
  \STATE \quad Update $\bm{\beta}_m^{(t+1)}$ by \textup{(\ref{eq:gamma_individual_beta_update})};
  \STATE \quad Update $\sigma_m^{2(t+1)}$ by \textup{(\ref{eq:gamma_individual_sigma_update})};
  \STATE \quad Update $\bm{\delta}_m^{(t+1)}$ by \textup{(\ref{eq:gamma_individual_delta_update})};
  \STATE \quad Update $\bm{u}_m^{(t+1)}$ by \textup{(\ref{eq:gamma_individual_dual_update})};
\ENDFOR
\ENSURE Each client $m \in [M]$ returns $\bm{\beta}_m^{(T)}$.
\end{algorithmic}
\end{algorithm}
For numerical stability, a lower bound of $10^{-12}$ was imposed on the
updated $\sigma_m^2$.
\subsection*{D.2\ \ \textit{$\gamma$-FedAvg}}
\subsubsection*{Optimization Problem}
In \textit{$\gamma$-FedAvg}, the following optimization problem is solved:
\begin{align}
  \mathop{\arg\min}_{\bm{\beta} \in \mathbb{R}^{p}} \; \frac{1}{M} \sum_{m=1}^{M}
  \Bigg\{ \frac{1}{2(1+\gamma)}\log\sigma_m^2
  + \frac{1}{2\sigma_m^2} \sum_{i=1}^{n_m} w_{mi}\, \bigl\{y_{mi}-\bm{x}_{mi}^\top\bm{\beta}\bigr\}^2 \Bigg\}
  +\sum_{j=1}^{p} p_{\lambda}(|\beta_{j}|).\notag
\end{align}
The first summation is derived from the loss function based on the
$\gamma$-divergence using the common regression coefficient $\bm{\beta}$
across all clients and the MM algorithm, whereas the second summation is a sparse
regularization term on the regression coefficients.
Therefore, \textit{$\gamma$-FedAvg} performs robust and sparse estimation of the common
regression coefficient $\bm{\beta}$ across all clients in the presence of data contamination.
\subsubsection*{Update Rules Based on ADMM}
To solve the optimization problem of \textit{$\gamma$-FedAvg}, we use ADMM.
To this end, we introduce an auxiliary variable
$\bm{\delta}\in\mathbb{R}^{p}$ and impose the constraint
$\bm{\beta}=\bm{\delta}$.
Here, we define
\begin{align*}
  f_{\gamma}(\bm{\beta}) &= \frac{1}{M} \sum_{m=1}^{M} \Bigg\{ \frac{1}{2(1+\gamma)}\log\sigma_m^2
  + \frac{1}{2\sigma_m^2} \sum_{i=1}^{n_m} w_{mi} \bigl\{ y_{mi}-\bm{x}_{mi}^\top\bm{\beta} \bigr\}^{2}
  \Bigg\}, \qquad h(\bm{\delta}) = \sum_{j=1}^{p} p_{\lambda}(|\delta_j|).
\end{align*}
With the above definitions, the optimization problem of \textit{$\gamma$-FedAvg}
can be reformulated as the following constrained optimization problem:
\begin{equation}
  \min \; f_{\gamma}(\bm{\beta}) + h(\bm{\delta}), \quad \text{s.t.} \quad \bm{\beta} = \bm{\delta}.\notag
\end{equation}
For this constrained optimization problem, the augmented Lagrangian function
$L_\rho(\bm{\beta},\bm{\delta},\bm{u})$ is defined, with Lagrange multiplier
$\bm{u}\in\mathbb{R}^{p}$ and penalty parameter
$\rho\in\mathbb{R}_{>0}$, as
\begin{align}
  L_{\rho}(\bm{\beta},\bm{\delta},\bm{u}) =&\, f_{\gamma}(\bm{\beta}) + h(\bm{\delta})
  + \bm{u}^\top (\bm{\beta}-\bm{\delta}) + \frac{\rho}{2} \| \bm{\beta}-\bm{\delta} \|_{2}^{2}.\notag
\end{align}
\paragraph{Update of $\bm{\beta}$:}
To efficiently update $\bm{\beta}$, we define a surrogate function
$F_\gamma(\bm{\beta})$ for $f_\gamma(\bm{\beta})$ as
$F_\gamma(\bm{\beta}) = f_\gamma(\bm{\beta}^{(t)})
+\allowbreak \nabla f_\gamma(\bm{\beta}^{(t)})^\top(\bm{\beta}-\bm{\beta}^{(t)})
+\allowbreak \frac{1}{2\tau}\|\bm{\beta}-\bm{\beta}^{(t)}\|_2^2$.
Here, $\tau>0$ is chosen to satisfy the upper-bound condition of the MM algorithm.
Therefore, we consider the following minimization problem using the surrogate function:
$\bm{\beta}^{(t+1)} = \mathop{\arg\min}_{\bm{\beta}}\bigl\{
F_\gamma(\bm{\beta}) +\allowbreak
\bm{u}^{(t)\top}(\bm{\beta}-\bm{\delta}^{(t)}) +\allowbreak
\frac{\rho}{2}\|\bm{\beta}-\bm{\delta}^{(t)}\|_2^2
\bigr\}$.
By differentiating this with respect to $\bm{\beta}$ and solving the corresponding stationary condition,
we obtain the following closed-form update:
\begin{align}
  \bm{\beta}^{(t+1)} = \frac{\tau}{1+\tau\rho} \left\{ \frac{1}{\tau}\bm{\beta}^{(t)}
  -\nabla f_\gamma(\bm{\beta}^{(t)}) +\rho\bm{\delta}^{(t)} -\bm{u}^{(t)} \right\}.
  \label{eq:gamma_fedavg_beta_update}
\end{align}
However, $\nabla f_\gamma(\bm{\beta}^{(t)})$ in
\eqref{eq:gamma_fedavg_beta_update} is obtained through server-side aggregation of the
local gradients computed by each client.
Here, we define the term corresponding to client $m$ in the summation of
$f_\gamma(\bm{\beta})$ as $f_{\gamma(m)}(\bm{\beta})$, such that
$f_\gamma(\bm{\beta}) = M^{-1}\sum_{m=1}^{M}f_{\gamma(m)}(\bm{\beta})$.
Accordingly, the following relation holds for the local gradients:
$\nabla f_\gamma(\bm{\beta}^{(t)})
= M^{-1}\sum_{m=1}^{M}\nabla f_{\gamma(m)}(\bm{\beta}^{(t)})$.
Here, the local gradient for each client $m$ is obtained as follows:
\begin{align}
  \nabla f_{\gamma(m)}(\bm{\beta}^{(t)}) &= \frac{1}{\sigma_m^{2(t+1)}}
  \left( \bm{X}_m^\top\bm{W}_m^{(t)}\bm{X}_m\bm{\beta}^{(t)} - \bm{X}_m^\top\bm{W}_m^{(t)}\bm{y}_m \right).
  \label{eq:gamma_fedavg_gradient}
\end{align}
Therefore, in the update step for $\bm{\beta}$, each client $m$
computes $\nabla f_{\gamma(m)}(\bm{\beta}^{(t)})$
using \eqref{eq:gamma_fedavg_gradient} and sends it to the server.
The server averages these local gradients to obtain
$\nabla f_\gamma(\bm{\beta}^{(t)})$ and updates the global model
$\bm{\beta}$ using \eqref{eq:gamma_fedavg_beta_update}.
\paragraph{Update of $\bm{\delta}$:}
Minimizing
$L_\rho(\bm{\beta}^{(t+1)},\bm{\delta},\bm{u}^{(t)})$
with respect to $\bm{\delta}$ gives
$\bm{\delta}^{(t+1)} = \mathop{\arg\min}_{\bm{\delta}}\bigl\{
h(\bm{\delta}) +\allowbreak \frac{\rho}{2}
\|\bm{\delta}-\bm{\beta}^{(t+1)}-\rho^{-1}\bm{u}^{(t)}\|_2^2
\bigr\}$.
Therefore, using the proximal operator $T_{\lambda,\rho}$,
the following update is obtained for each $j \in \{1, \dots, p\}$:
\begin{align}
  \delta_{j}^{(t+1)} = T_{\lambda,\rho} \left( \beta_{j}^{(t+1)} +\rho^{-1}u_{j}^{(t)} \right).
  \label{eq:gamma_fedavg_delta_update}
\end{align}

\paragraph{Update of $\bm{u}$:}
According to the update rule for the Lagrange multiplier in ADMM,
$\bm{u}$ is updated as follows:
\begin{align}
  \bm{u}^{(t+1)} = \bm{u}^{(t)} + \rho \bigl( \bm{\beta}^{(t+1)} - \bm{\delta}^{(t+1)} \bigr).
  \label{eq:gamma_fedavg_dual_update}
\end{align}

\paragraph{Update of $w_{mi}$:}
To ensure that the surrogate function obtained by applying the MM algorithm
to the loss function based on the $\gamma$-divergence satisfies the touching
condition at the current iterate, the weight $w_{mi}$ is updated as follows:
\begin{align}
  w_{mi}^{(t)} = \frac{ f_{y\mid x} \bigl(y_{mi}\mid \bm{x}_{mi};\,\bm{\theta}_m^{(t)}\bigr)^{\gamma} }{
  \sum_{k=1}^{n_m} f_{y\mid x} \bigl(y_{mk}\mid \bm{x}_{mk};\,\bm{\theta}_m^{(t)}\bigr)^{\gamma} }.
  \label{eq:gamma_fedavg_weight_update}
\end{align}
\paragraph{Update of $\sigma_m^2$:}
When minimizing with respect to $\sigma_m^2$,
the terms corresponding to the other clients are constant.
Therefore, the optimization problem can be solved independently for each client $m$ as
$\hat{\sigma}_m^2
= \mathop{\arg\min}_{\sigma_m^2}f_{\gamma(m)}(\bm{\beta},\sigma_m^2)$.
By differentiating
$f_{\gamma(m)}(\bm{\beta},\sigma_m^2)$
with respect to $\sigma_m^2$ and solving the corresponding stationary condition,
we obtain the following closed-form update:
\begin{align}
  \sigma_m^{2(t+1)} = (1+\gamma) \left( \bm{y}_m-\bm{X}_m\bm{\beta}^{(t)} \right)^\top \bm{W}_m^{(t)}
  \left( \bm{y}_m-\bm{X}_m\bm{\beta}^{(t)} \right).
  \label{eq:gamma_fedavg_sigma_update}
\end{align}
\subsubsection*{Optimization Algorithm}
In \textit{$\gamma$-FedAvg}, the weight $w_{mi}$ and variance $\sigma_m^2$,
which depend on the local data, are updated on each client.
Each client then computes the local gradient
$\nabla f_{\gamma(m)}(\bm{\beta}^{(t)})$
using the current global model $\bm{\beta}^{(t)}$
and sends it to the server.
The server computes the global gradient
$\nabla f_\gamma(\bm{\beta}^{(t)})$
by averaging the local gradients and updates the global model
$\bm{\beta}$, the auxiliary variable $\bm{\delta}$,
and the Lagrange multiplier $\bm{u}$,
which are shared across all clients.
Accordingly, the update scheme of \textit{$\gamma$-FedAvg} is summarized in
Algorithm~\ref{alg:gamma_fedavg}.
The initial values of $\bm{\beta}^{(0)}$ and $\sigma_m^{2(0)}$ are determined based on ridge regression, where $\bm{\beta}^{(0)}$ is obtained by averaging the client-specific ridge regression coefficients.
\begin{algorithm}[H]
\caption{\textit{$\gamma$-FedAvg} Algorithm}
\label{alg:gamma_fedavg}
\begin{algorithmic}[1]
\REQUIRE Tuning parameters $\tau, \rho, \gamma, \lambda$,
and $\iota$ only when SCAD regularization is used; number of iterations $T$;
\STATE \textbf{Initialization:}
Set $\bm{\delta}^{(0)}=\bm{0}$ and $\bm{u}^{(0)}=\bm{0}$;
Compute the client-specific ridge estimates
$\bm{\beta}_m^{\mathrm{Ridge}}$ and $\sigma_m^{2(0)}$ for each client $m$;
Set
$\bm{\beta}^{(0)} = M^{-1}\sum_{m=1}^{M}\bm{\beta}_m^{\mathrm{Ridge}}$;
\FOR{$t = 0, 1, \dots, T - 1$}
  \STATE \textbf{Server:}
  \STATE \quad Send $\bm{\beta}^{(t)}$ to the clients;
  \STATE \textbf{Clients $m \in [M]$ do in parallel:}
  \STATE \quad Update $w_{mi}^{(t)}$ for $i = 1, \dots, n_m$ by \textup{(\ref{eq:gamma_fedavg_weight_update})};
  \STATE \quad Update $\sigma_m^{2(t+1)}$ by \textup{(\ref{eq:gamma_fedavg_sigma_update})};
  \STATE \quad Calculate $\nabla f_{\gamma(m)}\!\left(\bm{\beta}^{(t)}\right)$
  by \textup{(\ref{eq:gamma_fedavg_gradient})};
  \STATE \quad Send $\nabla f_{\gamma(m)}\!\left(\bm{\beta}^{(t)}\right)$ to the server;
  \STATE \textbf{Server:}
  \STATE \quad Compute
  $\nabla f_\gamma(\bm{\beta}^{(t)})
  = M^{-1}\sum_{m=1}^{M}\nabla f_{\gamma(m)}(\bm{\beta}^{(t)})$;
  \STATE \quad Update $\bm{\beta}^{(t+1)}$ by \textup{(\ref{eq:gamma_fedavg_beta_update})};
  \STATE \quad Update $\bm{\delta}^{(t+1)}$ by \textup{(\ref{eq:gamma_fedavg_delta_update})};
  \STATE \quad Update $\bm{u}^{(t+1)}$ by \textup{(\ref{eq:gamma_fedavg_dual_update})};
\ENDFOR
\ENSURE Server returns $\bm{\beta}^{(T)}$.
\end{algorithmic}
\end{algorithm}
For numerical stability, a lower bound of $10^{-12}$ was applied to
$\sigma_m^2$ when computing the weights $w_{mi}$.

\end{document}